\documentclass[a4paper]{article}

\usepackage[margin=1in]{geometry} 

\usepackage{amsmath}
\usepackage{amsthm}
\usepackage{amssymb}

\usepackage[utf8]{inputenc}
\usepackage[hidelinks]{hyperref}
\hypersetup{
	unicode,
	pdfauthor={Author One, Author Two, Author Three},
	pdftitle={A simple article template},
	pdfsubject={A simple article template},
	pdfkeywords={article, template, simple},
	pdfproducer={LaTeX},
	pdfcreator={pdflatex}
}

\usepackage[sort&compress,numbers,square]{natbib}
\theoremstyle{plain}

\theoremstyle{definition}

\usepackage{graphicx, color}
\graphicspath{{fig/}}

\usepackage{color}

\usepackage{mathrsfs}

\def\be{\begin{equation}}
	\def\ee{\end{equation}}
\def\bse{\begin{subequations}}
	\def\ese{\end{subequations}}

\def\arccosh{\mathop{\rm arccosh}\nolimits}

\def\gl{\mathrel{\mathpalette\overl@ss>}}

\def\Real{\mathbb{R}}

\def\Complex{\mathbb{C}}

\def\Re{\mathop{\rm Re}\nolimits}
\def\Im{\mathop{\rm Im}\nolimits}

\def\arccosh{\mathop{\rm arccosh}\nolimits}

\def\@#1{{\mathbf{#1}}}
\def\_#1{{\mathsf{#1}}}
\let\~=\tilde
\let\==\bar
\let\^=\hat
\let\<=\langle
\let\>=\rangle

\def\min{\mathop{\rm min}\nolimits}

\def\note[#1]{\marginpar{\color{red}[#1]}}

\def\XXint#1#2#3{{\setbox0=\hbox{$#1{#2#3}{\int}$}
		\vcenter{\hbox{$#2#3$}}\kern-.5\wd0}}

\def\1{{\bf 1}}

\makeatother
\usepackage{color}

\title{On the discrete Kuznetsov–Ma solutions for the defocusing Ablowitz-Ladik equation with large background amplitude}
\author{E. C. Boadi$^1$, E. G. Charalampidis$^{2,3}$, P. G. Kevrekidis$^4$, N. J. Ossi$^{1}$, B. Prinari$^{1}$}

\date{$^1$ Department of Mathematics, State University of New York, Buffalo, NY 14260\\
$^2$ Department of Mathematics and Statistics, San Diego State University, San Diego, CA 92182\\
$^3$ Computational Science Research Center, San Diego State University, San Diego, CA 92182\\
$^4$ Department of Mathematics and Statistics, University of Massachusetts, Amherst, MA 01003}

\begin{document}
	\maketitle
	\begin{abstract}
The focus of this work is on a class of solutions of the defocusing Ablowitz-Ladik lattice on an arbitrarily large background which are discrete analogs of the
Kuznetsov-Ma (KM) breathers of the focusing nonlinear Schr\"odinger equation.
One such solution was obtained in 2019 as a byproduct of the Inverse Scattering Transform, and it was observed that the solution could be regular for certain choices of the soliton parameters, but its regularity was not analyzed in detail. This work provides a systematic investigation of the conditions on the background and on the spectral parameters that guarantee the KM solution to be non-singular on the lattice for all times. Furthermore, a novel KM-type breather solution is presented which is also regular on the lattice under the same conditions. We also employ Darboux transformations to obtain a multi-KM breather solution, and show that parameters choices exist for which a double KM breather solution is regular on the lattice.
We analyze the features of these solutions, including their
frequency which, when tending to 0, renders them proximal to
rogue waveforms.
Finally, numerical results on the  stability and
spatio-temporal dynamics of the single KM breathers are presented, showcasing
the potential destabilization of the obtained states due to the
modulational instability of their background.
	\end{abstract}
	\section{Introduction}

The Ablowitz-Ladik (AL) model:
\begin{equation}
\label{e:ALq}
i\frac{dq_n}{dt}=\frac{1}{h^2}(q_{n+1}-2q_n+q_{n-1})-\sigma |q_n|^2(q_{n+1}+q_{n-1}), \quad \sigma=\pm 1, \quad n\in \mathbb{Z},t\in\Real
\end{equation}
was introduced in \cite{AL1,AL2} as an integrable spatial discretization of the focusing ($\sigma=-1$) and defocusing ($\sigma=1$) nonlinear Schr\"odinger (NLS) equation:
$$
iq_t=q_{xx}-2\sigma |q|^2q,
$$
to which they reduce as the lattice spacing $h\to 0$ with $nh\to x$.

Besides being used as numerical schemes for their continuous counterparts, the AL equations describe the quantum correlation function of the $XY$-model of spins \cite{IIKS1991,LSM1961}. When $\sigma = -1$, the equation plays a significant role in the study of anharmonic lattices \cite{TH1990} and lossless nonlinear electric lattices \cite{MBR1995}, and it is gauge equivalent to an integrable discretization of the Heisenberg spin chain \cite{I1982}. Additionally, the focusing AL hierarchy describes the integrable motions of a discrete curve on the sphere \cite{DS1995}.
Importantly, in addition to its intrinsic value, the AL model
provides a springboard towards exploring solutions in the non-integrable,
yet more widespread discrete variant of the nonlinear Schr{\"o}dinger
model, namely the so called discrete NLS equation~\cite{dnlsbook}.
Indeed, this has been leveraged for solitary wave dynamics in
early works, such as, e.g.,~\cite{cai}, but also for more
complex structures including periodic and rogue waves, e.g.,
in the more recent work of~\cite{SCCKK2020}. This justifies a
multi-faceted interest in the periodic (and rogue) wave
solutions of the Ablowitz-Ladik model~\cite{AASC2009}, from
a theoretical point of view in their own right, but also from
a computational and applications point of view, as well.

In this work we are interested in characterizing special solutions of the defocusing AL equation on a nontrivial background. Specifically, we will consider the equation:
\begin{equation}
\label{e:AL}
i\frac{dQ_n}{d\tau}=Q_{n+1}+Q_{n-1}+2r^2Q_n-|Q_n|^2(Q_{n+1}+Q_{n-1}), \qquad r^2=Q_o^2-1>0
\end{equation}
with boundary conditions:
\begin{equation}
\label{e:BG}
\lim_{n\to \pm \infty}Q_n(\tau)=Q_{\pm}\equiv Q_o e^{i\theta_\pm}\qquad Q_o>1,\theta_\pm\in \Real.
\end{equation}
Modulo a rescaling of dependent and independent variables ($Q_n= hq_n$ and $\tau=t/h^{-2}$), the above equation is gauge-equivalent to Eq.~\eqref{e:ALq} (the gauge transformation
$Q_n\to Q_n=e^{-2iQ_o^2\tau}$ has simply the effect of allowing for constant boundary conditions as $n\to \pm \infty$).

As mentioned above, the AL equations are integrable, and their initial-value problem can be solved by means of the Inverse Scattering Transform (IST).
The IST for Eq.~\eqref{e:AL} was initially developed under the assumption that the background amplitude $Q_o$ satisfies
a ``small norm'' condition, $0<Q_o<1$ \cite{VK1992,CKVV1992,ABP2007}.
This condition is actually not just a technical assumption: in \cite{OY2014} it was shown that the defocusing AL equation, which is modulationally stable for $0\le Q_o<1$,
becomes unstable if $Q_o>1$, and exhibits discrete rogue wave solutions, some of which are regular for all times.
This finding sparked renewed interest in the defocusing AL equation on a background. In \cite{OP2019} the IST was generalized to the case of arbitrarily large background $Q_o>1$.
As a by-product of the IST, explicit solutions were also obtained in \cite{OP2019}, which are the discrete
analog of the celebrated Kuznetsov–Ma \cite{KM1997,Ma1979}, Akhmediev \cite{AEK1985,AK1987}, Tajiri-Watanabe \cite{TW1998} and Peregrine solutions \cite{Per1983,AAT2009,AASC2009}, and which mimic the corresponding solutions for the focusing AL equation \cite{AASC2010,ADUCA2013,PV2016,P2016}.

The recent work \cite{CS2024} further investigated the stability of background solutions of the AL equations in the
periodic setting, confirming in particular that, unlike its continuous counterpart, in the defocusing case ($AL_-$, in the terminology of \cite{CS2024}) the background solution
is unstable under a monochromatic perturbation of any wave number if the amplitude of the background is greater than 1.
A discrete Kuznetsov-Ma (KM) breather solution was presented in \cite{OP2019}, and it was observed that the solution could be regular for certain choices of the soliton parameters, but its regularity was not analyzed in detail.

The goal of the present work is to provide a systematic investigation of the conditions on the background and on the spectral parameters (discrete eigenvalue and associated norming constant) that guarantee the KM solution to be non-singular for all $n\in \mathbb{Z}$ and all $\tau\in \Real$. Furthermore, a novel KM-type breather solution will be presented which is also regular on the lattice under the same conditions. We also employ the Darboux transformations derived in \cite{CS2024} to analyze Akhmediev breathers, and, by generalizing the construction to the case of KM breathers, we obtain multi-KM breather solutions. In particular, we show that
parameters choices exist for which a double KM breather is regular on the lattice.
Finally, we corroborate the analytical findings by numerical
computations where for the obtained exact solutions, a spectral
stability analysis is performed and dynamical evolution of the
lattice model with such waveforms as initial data is explored.
The latter will often lead to unbounded growth of the solution
due to the manifestation of the modulational instability of its background.

Our presentation is structured as follows. First, we provide
some relevant background of the model, including
its Lax pair and associated discrete spectrum. We then move to
the construction of the KM breathers, followed by an analysis of the
conditions under which the relevant waveforms are regular.
We then proceed to provide a construction of the solutions
via Darboux transformations, before exploring the stability and
direct numerical dynamics of the associated waveforms. We close
with a discussion of the conclusions of our findings and some
directions for future research.

\section{Lax pair and discrete spectrum}
The Lax pair for the AL system \eqref{e:AL} is given by
\cite{AL1}:
\bse
\begin{gather}
\label{Lax}
v_{n+1}=\mathbf{L}_n v_n\,, \qquad
\frac{d v_n}{d \tau} = \mathbf{M}_n v_n\, , \\
\mathbf{L}_n=\mathbf{Z}+\mathbf{Q}_n\,,\qquad
\mathbf{Z}=\left(\begin{array}{cc}z & 0 \\ 0 & 1/z\end{array}\right)\,, \qquad
\mathbf{Q}_n=\left( \begin{array}{cc}0 & Q_n \\ Q_n^*
& 0
\end{array}\right)\,, \label{Q,Z} \\
\mathbf{M}_n=i\sigma_3 \left[\mathbf{Q}_n\mathbf{Q}_{n-1} - Q_o^2\mathbf{I}-\frac{1}{2} (z-1/z)^2 \mathbf{I}-\mathbf{Q}_n\mathbf{Z}^{-1}+\mathbf{Q}_{n-1}\mathbf{Z}\right]\,,
\end{gather}
\ese
where $v_n$ is a $2$-component vector, $z\in \mathbb{C}$ is the
scattering parameter, and $Q_n(\tau)$ is the AL potential.

If $Q_n(\tau)$ satisfies \eqref{e:BG} as $n\to\pm \infty$, the asymptotic behavior of the simultaneous solutions of the Lax
pair is given by $\lambda^n(ir)^n e^{i\Omega \tau}$ or $\lambda^{-n}(ir)^{-n}e^{-i\Omega \tau}$, where
$\lambda$ as a function of $z$ is defined by
\begin{equation}
ir(\lambda+1/\lambda)=z+1/z\,, \qquad r =\sqrt{Q_o^2-1}\in \Real^+\,, \label{lambda}
\end{equation}
and the dispersion relation is given by:
\begin{equation}
\Omega(z)=-\frac{i}{2}r(z-1/z)(\lambda-1/\lambda)\equiv -\frac{ir}{2\lambda z}(z^2-1)(\lambda^2-1)\,. \label{e:Omega}
\end{equation}
Clearly, $\lambda=\lambda(z)$ has branching, with 4 branch points on the imaginary axis located at:
\begin{equation}
z=i(r\pm Q_o)\,, \qquad z=i(-r\pm Q_o)\, .\label{branch_points}
\end{equation}
Note that $-r-Q_o<-1<r-Q_o<0<Q_o-r<1<Q_o+r$, which specifies the order of the branch points.
\begin{figure}[t!]
\includegraphics[height=5cm]{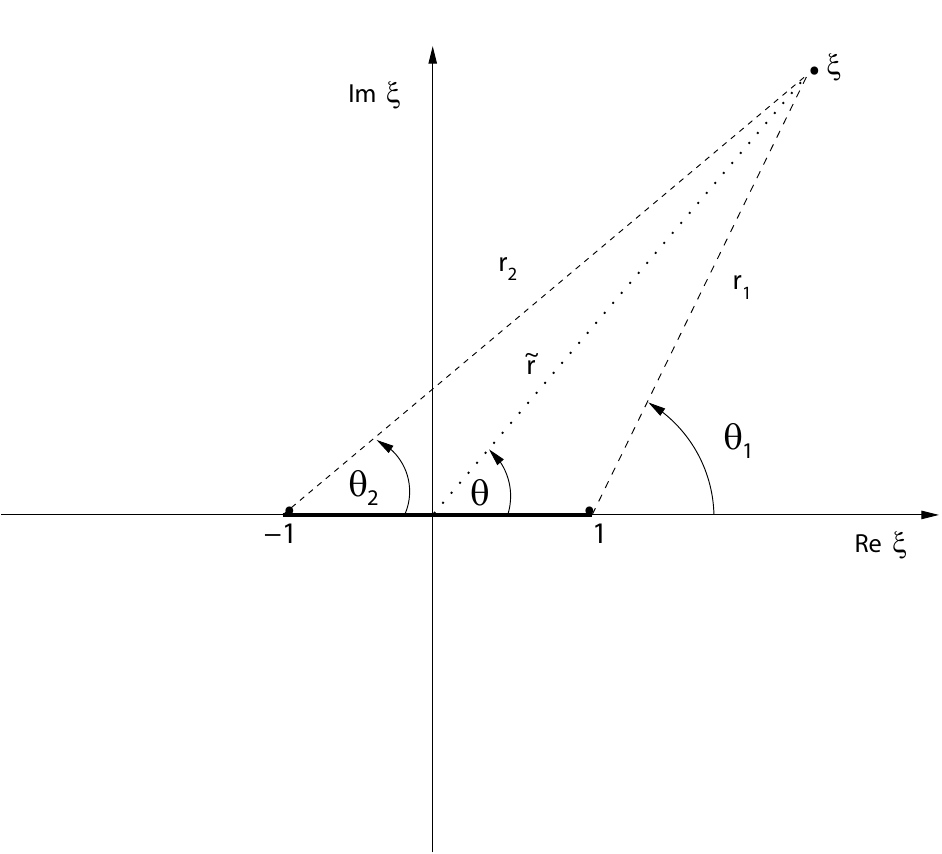}\qquad \qquad
\includegraphics[height=5cm]{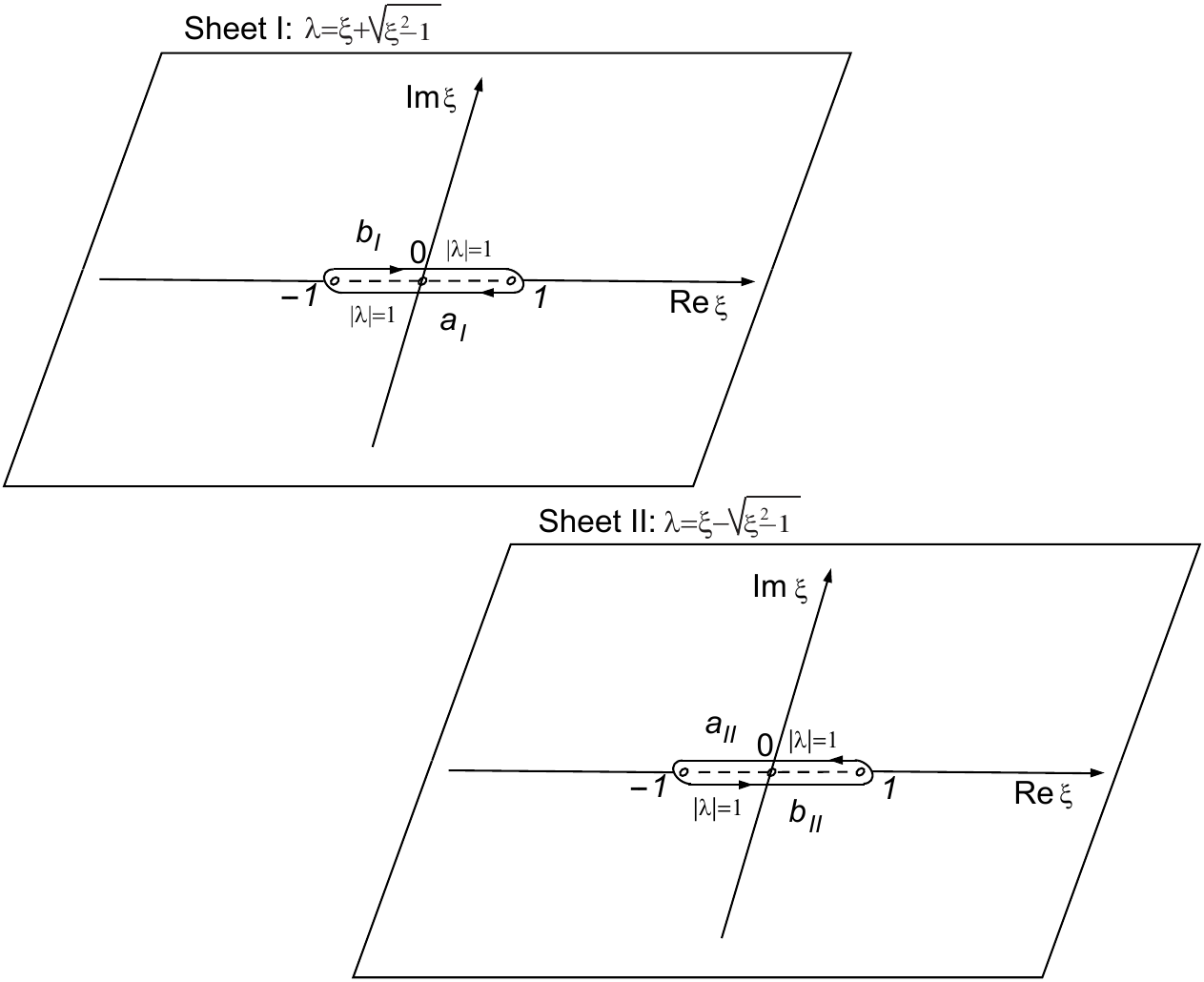}
\caption{Left: The choice of branch cut in the complex $\xi$-plane, $\xi=(z+1/z)/2ir$.
Right: The 
2-sheeted covering defined by
$\lambda=\xi+(\xi^2-1)^{1/2}$.} \label{f:fig-branch_xi}
\end{figure}
It is convenient to introduce the variable $\xi=(z+1/z)/2ir$, and define a 2-sheeted Riemann surface cut across the segment $(-1,1)$ on the real $\xi$-axis,
with local coordinates on each sheet given as in the left panel of Fig.~\ref{f:fig-branch_xi}:
\begin{align}
& \xi-1=r_1e^{i\theta_1}\, , \qquad \xi+1=r_2e^{i\theta_2} \qquad 0\le\theta_1,\theta_2 <2 \pi \label{T1,2}\\
& \xi =\tilde{r}e^{i\theta} \qquad 0\le\theta <2 \pi
\end{align}
such that on Sheet I/II $\lambda(z)=\xi \pm \sqrt{\xi^2-1}$.
The IST developed in \cite{OP2019} is formulated in terms of a uniformization variable
\begin{equation}
\zeta(z)=\lambda(z)/z\, , \label{zeta}
\end{equation}
for which one has
\begin{equation}
z^2=\frac{\zeta-ir}{\zeta( ir\zeta-1)
},\qquad\lambda^2 =\zeta\frac{\zeta-ir}{ir\zeta-1},\qquad
z\lambda=\frac{\zeta-ir}{ir\zeta-1} \label{z^2_etc}\,,
\end{equation}
showing that all these quantities, as well as all quantities which are even
functions of $\lambda$ and $z$,
are meromorphic functions of $\zeta$. It is worth noticing that $|\lambda^2(\zeta)|\lessgtr 1$ for $\zeta\in D_\pm$, where $D_\pm=\left\{\zeta\in\Complex:\, |\zeta-ir|\lessgtr Q_o \right\}$, and $|\lambda(\zeta)|^2=1$ for $\zeta\in \Sigma$, where $\Sigma=\left\{\zeta\in\Complex:\, |\zeta-ir|=Q_o \right\}$.
We refer the reader to \cite{OP2019} for the details about the IST, while noting that in Appendix~B we provide a short errata to correct some typos and small inaccuracies in \cite{OP2019}. For the purposes of this work, it suffices to recall that the solitons/breathers of the AL equation are characterized by special values of the spectral parameter ($z$, or $\zeta$) for which bounded eigenfunctions of the Lax pair exist. In terms of the uniformization variable $\zeta$ such discrete eigenvalues appear in symmetric quartets:
$
\left\{ \zeta_k,\, \frac{\zeta_k^*+ir}{ir\zeta_k^*+1},\, -1/\zeta_k^*,\,
\frac{ir\zeta_k-1}{\zeta_k-ir} \right\}
$,
with the first two in $D_-$ and the last two in $D_+$. It is convenient to then group the eigenvalues as follows:
\begin{equation}
\label{e:set_discrete_eigenvalues}
\left\{ \zeta_k, \bar{\zeta}_k\right\}_{k=1}^{2J}\,, \qquad \zeta_k=\frac{\zeta_{k-J}^*+ir}{ir\zeta_{k-J}^*+1} \quad \text{for } k=J+1,\dots,2J
\end{equation}
and $\bar{\zeta}_k=-1/\zeta_k^*$ for $k=1,\dots,2J$. As is typical, the discrete eigenvalue specifies the velocity and amplitude of the corresponding soliton/breather.
To each discrete eigenvalue $\bar{\zeta}_k\in D_+$ is associated a norming constant $\bar{C}_k$:
\begin{equation}
C_k=-\left(\bar{C}_k/\bar{\zeta}_k^2\right)^*,\qquad C_{k+J}=-\frac{(Q_+^*)^2}{(\bar{\zeta}_k-ir)^2}\bar{C}_k\quad k=1,\dots,J
\end{equation}
which in turn characterizes the soliton/breather position and phase.

The general discrete breather solution of Eq.~\eqref{e:AL} on the nontrivial background \eqref{e:BG} is given in Eq.~(89) in \cite{OP2019},
and it is an analog of the discrete Tajiri-Watanabe (TW) breather of the focusing AL equation obtained in \cite{P2016}. The solution is given in terms of a  discrete eigenvalue $\bar{\zeta}_1\in D_+$ and its associated norming constant $\bar{C}_1$:
\begin{equation}
\label{e:TW}
Q_n(\tau)=Q_+\left\{ 1+\frac{ir}{2Q_+^*\sqrt{\delta_4}}\frac{(\delta_1^*+\delta_2^*)\cos \eta-i(\delta_1^*-\delta_2^*)\sin \eta+\delta_3^* e^{-\xi}}{\cosh{(\xi-\xi_o)}+\alpha\cos(\eta-\eta_o)} \right\},
\end{equation}
where
\bse
\label{e:TWcoeffs}
\begin{gather}
\delta_1\pm \delta_2=\bar{C}_1\frac{\bar{\zeta}_1-ir}{1-ir\bar{\zeta}_1}\pm \bar{C}_1^*\frac{Q_+^2\bar{\zeta}_1^*}{(1+ir\bar{\zeta}_1^*)^2}\,, \qquad
\delta_3=-\frac{|\gamma|^2(\bar{\zeta}_1-ir)|\bar{\zeta}_2-\bar{\zeta}_1|^2(\bar{\zeta}_1+\bar{\zeta}_1^*)}{Q_+^*(\zeta_1-ir)(1+|\bar{\zeta}_1|^2)|\bar{\lambda}_1^2-1|^2}\,,\\
\delta_4=\frac{|\gamma|^2|\bar{\zeta}_1|^2|\bar{\zeta}_1-ir|^2(|1-ir\bar{\zeta}_1|^2-Q_o^2)^2}{r^2Q_o^2|\bar{\lambda}_1^2-1|^2(1+|\bar{\zeta}_1|^2)^2|1-ir\bar{\zeta}_1|^2}\,, \qquad
\gamma=\frac{Q_+^*\bar{C}_1}{1-ir\bar{\zeta}_1}\,,
\qquad
\tilde{A}_1+\gamma=\frac{\gamma \bar{\lambda}_1^2}{\bar{\lambda}_1^2-1}\,, \\
\xi=n\log \rho+2(\Im \bar{\Omega}_1)\tau\,, \qquad \eta=n\phi +2(\Re \bar{\Omega}_1)\tau, \\
\alpha=|\tilde{A}_1+\gamma|/\sqrt{\delta_4}>1\,,\\
\xi_o=\log \sqrt{\delta_4}\,, \qquad 
\qquad \cos \eta_o= \frac{\Re (\tilde{A}_1+\gamma)}{|\tilde{A}_1+\gamma|}\,, \qquad
\sin \eta_o=-\frac{\Im (\tilde{A}_1+\gamma)}{|\tilde{A}_1+\gamma|}\,, \\
\bar{\Omega}_1\equiv \Omega(\bar{\zeta}_1)=\frac{r^2(\bar{\zeta}_1^2+2i\bar{\zeta}_1/r+1)(\bar{\zeta}_1^2-2ir\bar{\zeta}_1+1)}{2\bar{\zeta}_1(\bar{\zeta}_1-ir)(1-ir\bar{\zeta}_1)}.
\end{gather}
\ese
According to \eqref{z^2_etc}, $\lambda^2(\zeta)=\zeta(\zeta-ir)/(ir\zeta-1)$ and
\begin{equation}
\label{e:rho}
\bar{\lambda}_1^2=\lambda^2(\bar{\zeta}_1)
=:e^{i\phi}/\rho.
\end{equation}
Also, $\bar{\zeta}_2=(\bar{\zeta}_1^*+ir)/(ir\bar{\zeta}_1^*+1)$ and $\zeta_1=-1/\bar{\zeta}_1^*$, and we recall that $|\lambda^2(\zeta)|< 1$ for any
$\zeta\in D_+$ (hence, $\rho>1$). It is also important to point out that $\alpha>1$ for any choice of the soliton parameters $\bar{\zeta}_1$ and $\bar{C}_1$, which means that
in general the TW solution is singular.
This singularity, as well as ways to avoid it, is crucial for our considerations
that follow next.
We refer the reader to \cite{OP2019} for the details of the derivation of Eq.~\eqref{e:TW}, while noting that we corrected a typo in the expressions for $\xi,\eta$ (see also Appendix~B).

\section{Kuznetsov-Ma (KM) breathers}
The discrete analog of the Kuznetsov-Ma (KM) solutions is obtained when the discrete eigenvalue $\bar{\zeta}_1\in D_+\setminus\left\{0,r\right\}$ is purely imaginary, i.e.,
$\bar{\zeta}_1=i\kappa$ with
$$
\kappa\in (-\sqrt{1+r^2}+r,\sqrt{1+r^2}+r)\setminus \left\{0,r\right\},
$$
where we recall $r=\sqrt{Q_o^2-1}>0$.
In this case, the parameter $\bar{\lambda}_1^2=\lambda^2(\bar{\zeta}_1)\in \Real$, but note that it is not necessarily positive.
Specifically, one has
\begin{equation}
\bar{\lambda}_1^2=\frac{\kappa (\kappa-r)}{1+r\kappa}
\end{equation}
and one can check that  $-1/r<-\sqrt{1+r^2}+r$ for any $r$, therefore
\begin{align*}
& \bar{\lambda}_1^2>0 \quad \text{for} \quad -\sqrt{r^2+1}+r<\kappa<0 \ \text{ and } \ r<\kappa<\sqrt{r^2+1}+r \\
& \bar{\lambda}_1^2<0 \quad \text{for} \quad 0<\kappa<r\,.
\end{align*}
In \cite{OP2019} the implicit assumption was made that $\bar{\lambda}_1^2>0$, but in the parametrization $\bar{\lambda}_1^2=e^{i\phi}/\rho$ one needs to take $\phi=\pi$ if $\bar{\lambda}_1^2<0$, while indeed $\phi=0$ when $\bar{\lambda}_1^2>0$. This will not affect the regularity of the solution (see below), but it will give rise to a different solution if one chooses values of $r,\kappa$ for which $\bar{\lambda}_1^2<0$.
Based on the above discussion, we have:
\begin{equation}
\label{e:rhok}
\rho=\left| \frac{1+\kappa r}{\kappa(\kappa-r)}\right|>1\,,
\end{equation}
and
\begin{gather}
\phi=0 \qquad \text{for } -\sqrt{1+r^2}+r<\kappa<0 \quad \text{or} \quad r<\kappa<\sqrt{r^2+1}+r\\
\phi=\pi \qquad \text{for} \qquad 0<\kappa<r.
\end{gather}
Now, considering \eqref{e:TWcoeffs} when $\bar{\zeta}_1=i\kappa$, one can immediately see that $\delta_3=0$ and
\begin{equation}
\bar{\Omega}_1=\frac{r(r\kappa^2+2\kappa -r)(-\kappa^2+2r\kappa +1)}{2\kappa(\kappa-r)(1+\kappa r)}\in \Real \qquad \text{[corrected a typo in Eq.~(94) in \cite{OP2019}]}\,.
\end{equation}
The remaining coefficients \eqref{e:TWcoeffs} in the TW solution reduce to
\bse
\begin{align}
\delta_4&= \frac{|\bar{C}_1|^2\kappa^2(\kappa-r)^2(r\kappa^2+2\kappa-r)^2}{(1+\kappa^2)^2(1+\kappa r)^2(\kappa^2-2\kappa r-1)^2}>0,
\\
\delta_1\pm \delta_2&=i\bar{C}_1\frac{\kappa-r}{1+\kappa r}\mp i\bar{C}_1^* Q_+^2\frac{\kappa}{(1+\kappa r)^2},
\\
\tilde{A}_1+\gamma&=\bar{C}_1\frac{Q_+^*\kappa(\kappa -r)}{(1+r\kappa)(\kappa^2-2r\kappa -1)},
\end{align}
\ese
and
\bse
\begin{gather}
\xi=n\log \rho\,, \qquad \eta=n \phi +2\bar{\Omega}_1\tau\,, \qquad \xi_o=\log \sqrt{\delta_4}, \label{e:xi,eta,xio} \\
\alpha=\frac{|\tilde{A}_1+\gamma|}{\sqrt{\delta_4}}\equiv\frac{(1+\kappa^2)\sqrt{1+r^2}}{|r\kappa^2+2\kappa  -r|}>1, 
\label{e:KMalpha}
\\
\cos \eta_o =\frac{\Re (\tilde{A}_1+\gamma)}{|\tilde{A}_1+\gamma|}\equiv -\mathrm{sgn}(\bar{\lambda}_1^2) \cos(\chi-\theta)\,, \quad
\sin \eta_o =-\frac{\Im (\tilde{A}_1+\gamma)}{|\tilde{A}_1+\gamma|}\equiv \mathrm{sgn}(\bar{\lambda}_1^2) \sin(\chi-\theta)\,, 
\end{gather}
\ese
where we have written the norming constant $\bar{C}_1=c\, e^{i\chi}$, $c>0$, and the background as $Q_+=Q_o\, e^{i\theta}\equiv e^{i\theta}\sqrt{1+r^2}$, and to simplify
the expressions of $\cos \eta_o,\sin\eta_o$ we have taken into account that $\kappa^2-2r\kappa-1<0$ for $\kappa$ in the allowed range, and that the remaining factor in $\tilde{A}_1+\gamma$ is simply $\bar{\lambda}_1^2$. Note the additional term $n\phi$ in the expression \eqref{e:xi,eta,xio} of $\eta$ compared to \cite{OP2019}, which, we reiterate, is only present if $0<\kappa<r$.

\begin{figure}[h!]
\includegraphics[width=\textwidth]{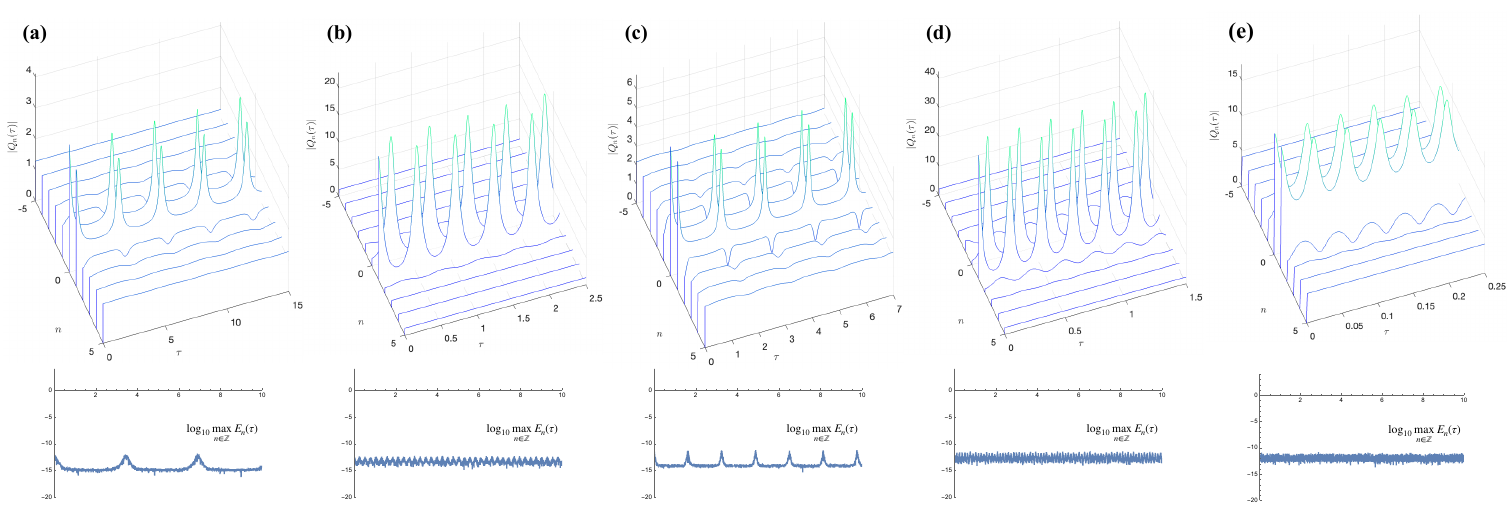}
\centering
\caption{Regular Kuznetsov-Ma solutions corresponding to \textbf{(a)} $r=0.8$, $\kappa=-0.3$, c=1.23 (Period: 3.44946); \textbf{(b)} $r=0.8$, $\kappa=0.7$, $c=257.5$ (Period: 0.530412); \textbf{(c)} $r=2$, $\kappa=3.5$, $c=2.32339$ (Period: 1.62646); \textbf{(d)} $r=2$, $\kappa=0.09$, $c=13.5884$ (Period: 0.261321); \textbf{(e)} $r=4$, $\kappa=4.5$, $c=101.562$ (Period: 0.0466168). \textbf{(a)}, \textbf{(c)}, and \textbf{(e)} are KM1 solutions, i.e., given by \eqref{e:KM1}, while \textbf{(b)} and \textbf{(d)} are KM2 solutions, i.e., corresponding to \eqref{e:KM2}. To confirm that the solutions presented are correct, below each panel is a log-plot of the error maximized over $n\in\mathbb{Z}$ as a function of $\tau$.
}
\label{1KM_Figure}
\end{figure}

For simplicity, we can assume below that $\theta_+=0$, i.e., $Q_+=Q_o$, and also take the norming constant to be real and positive, i.e., $\chi=0$ and $\bar{C}_1=c>0$.
Then, if $r,\kappa$ are chosen so that $\bar{\lambda}_1^2>0$, $\eta_o=\pi$ and
from \eqref{e:TW} we obtain
\begin{equation}
\label{e:KM1}
Q_n(\tau)=Q_o+\frac{ir}{2\sqrt{\delta_4}}\frac{\left(\delta_1^*+\delta_2^*\right)\cos{(2\bar{\Omega}_1\tau)}- i \left(\delta_1^*-\delta_2^*\right)\sin{(2\bar{\Omega}_1\tau)}
}{\cosh (n\log \rho-\xi_o)-\alpha \cos{(2\bar{\Omega}_1\tau)}}\,.
\end{equation}
Conversely, still assuming that the phases of $Q_+$ and $\bar{C}_1$ are zero, if $r,\kappa$ are chosen so that $\bar{\lambda}_1^2<0$, then
$\eta_o=0$, and \eqref{e:TW} yields
\begin{equation}
\label{e:KM2}
Q_n(\tau)=Q_o+\frac{ir(-1)^n}{2\sqrt{\delta_4}}\frac{\left(\delta_1^*+\delta_2^*\right)\cos{(2\bar{\Omega}_1\tau)}- i \left(\delta_1^*-\delta_2^*\right)\sin{(2\bar{\Omega}_1\tau)}
}{\cosh (n\log \rho-\xi_o)+(-1)^n\alpha \cos{(2\bar{\Omega}_1\tau)}}
\,.
\end{equation}
Eq.~\eqref{e:KM1} was already obtained in \cite{OP2019}, and we only corrected a typo. In the remainder of this work, we will refer to this solution as KM1.
On the other hand, to the best of our knowledge Eq.~\eqref{e:KM2} is a novel KM breather solution of the defocusing AL equation with background $Q_o>1$, which we will refer to as KM2.
Some prototypical case examples
of both families of solutions are shown in Fig.~\ref{1KM_Figure}.
There, it can be seen that the examples of KM2 breathers, carrying the
staggered spatial structure evident in Eq.~(\ref{e:KM2}), feature
peaks that bear an alternating structure in time, contrary
to what is observed for the KM1 breathers.

\section{Regularity conditions for the KM breathers}
Note that $\alpha>1 $ for any $\kappa\ne r$ [see Eq.~\eqref{e:KMalpha}], which means that the denominators in Eq.~\eqref{e:KM1} and Eq.~\eqref{e:KM2} will generically have zeros, and hence the solution will be singular unless, for a given $r$ (specified by $Q_o$), one can choose the soliton parameters $\kappa$ and $\bar{C}_1$ so that no integer falls in the interval where $\cosh(n\log \rho-\xi_o)\le \alpha$.

Indeed, the minimum value for the $\cosh$ term in the denominator is 1, and it is achieved when $x=\xi_o/\log \rho$, where $\xi_o=\log \sqrt{\delta_4}$. The interval where $\cosh$ is less than or equal to $\alpha$ is given by
\begin{equation}
\label{e:interval}
(\xi_o-\arccosh \alpha )/\log\rho<x<(\xi_o+\arccosh \alpha)/\log \rho,
\end{equation}
and the width of this interval is $2\arccosh \alpha/\log \rho$, which is independent of the norming constant $\bar{C}_1$, while the center of the interval, which is the center of the soliton, $\xi_o/\log \rho$, is determined by $|\bar{C}_1|$ via $\delta_4$.

\begin{figure}[h!]
\includegraphics[width=0.9\textwidth]{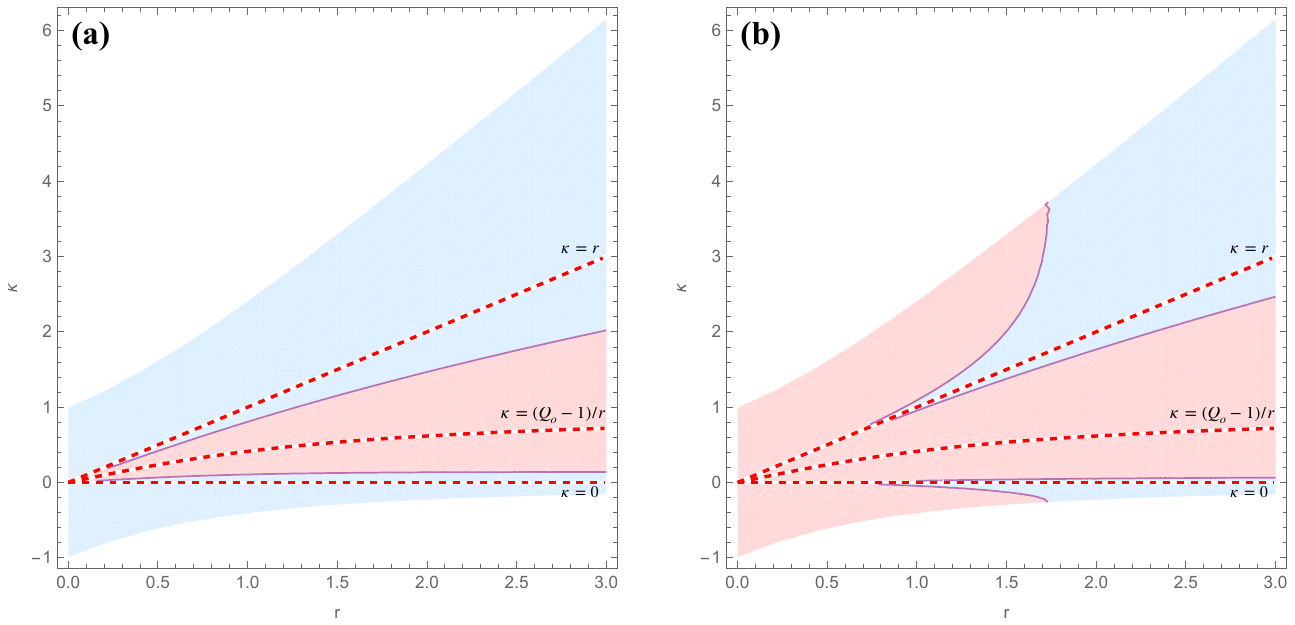}
\centering
\caption{\textbf{(a)} Contour plot of $f(r,\kappa)=\arccosh \alpha-\log \sqrt{\rho}$ as a function of $r$ (horizontal axis) and $\kappa$ (vertical axis) for $r-\sqrt{r^2+1}<\kappa<r+\sqrt{r^2+1}$. The blue shaded regions correspond to $f(r,\kappa)<0$ (where the regularity condition is satisfied) and the red shaded regions correspond to $f(r,\kappa)>0$ (where the regularity condition is not satisfied). The dashed red lines correspond to $\kappa=0,(Q_o-1)/r,r$, where $f(r,\kappa)$ is singular. Note that if $r-\sqrt{r^{2}+1}<\kappa<0$ or $r<\kappa<r+\sqrt{r^2+1}$ the solution is a KM1, while if $0<\kappa<r$ the solution is a KM2. \textbf{(b)} Same for $g(r,\kappa)=\arccosh \alpha-\log (\rho)^{1/4}$. In order for the stronger regularity condition to be satisfied, $(r,\kappa)$ should be chosen in the blue regions.}
\label{f:contour}
\end{figure}

A necessary condition in order for the above interval \eqref{e:interval} not to contain any integer is that the width of the interval be strictly less than 1, and hence a necessary condition for regularity of the solution is that
\begin{equation}
\label{e:reg_cond}
f(r,\kappa):=\arccosh\left[ \frac{(1+\kappa^2)\sqrt{1+r^2}}{|r\kappa^2+2\kappa  -r|}\right]
-\frac{1}{2}\log \left|\frac{1+\kappa r}{\kappa(\kappa-r)}\right|<0.
\end{equation}
The latter is an inequality for the two parameters $r,\kappa$, and one can find pairs of values $r_o,\kappa_o$ with $r_o>0$ and $-\sqrt{r_o^2+1}+r<\kappa_o<\sqrt{r_o^2+1}+r_o$,
$\kappa_o\ne 0,r_o$ for which this necessary condition for regularity is satisfied; see, in particular, the representation of the left panel of Fig.~\ref{f:contour}. Note that the set of loci of $(r,\kappa)$ for which \eqref{e:reg_cond} is satisfied is certainly not empty, since $\log \rho$ in the second term is positive (recall $\rho>1$), and it becomes arbitrarily large as $\kappa \to r$ from either side. The contour plot of $f(r,\kappa)$ is shown on the left panel in Fig.~\ref{f:contour}. A stronger regularity condition, i.e., requiring the width of the interval where $\cosh(n\log \rho -\xi_o)-\alpha\le 0$ to be smaller than 1/2, can be imposed to ensure a larger lower bound on the denominator of the solution. The corresponding contour plot is shown on the right panel in Fig.~\ref{f:contour}.

Once $r_o,\kappa_o$ are chosen to satisfy \eqref{e:reg_cond} so that the interval has width less than 1, we need to choose the center of the interval so that it does not contain any integer. We can fix $\bar{n}\in \mathbb{Z}$
arbitrarily, and place the center the soliton in the interval $(\bar{n},\bar{n}+1)$ by requiring:
\begin{equation}
\label{e:center_cond}
\bar{n}+\Delta<\frac{\xi_o}{\log \rho}<\bar{n}+1-\Delta,
\end{equation}
where the half width of the interval, $\Delta$, is
\begin{equation}
\label{e:Delta}
\Delta=\frac{\arccosh \alpha}{\log \rho} = \frac{\arccosh\left[ (1+\kappa^2)\sqrt{1+r^2}|r\kappa^2+2\kappa  -r|^{-1}\right]}{\log \left|\frac{1+\kappa r}{\kappa(\kappa-r)}\right|} <\frac{1}{2}.
\end{equation}
[Note the width of the above interval is exactly $1-2\Delta>0$.] The smaller the value of $\Delta$, the further away from the singularities one
can choose the closest integers, and therefore the smaller the value of $Q_n(\tau)$.

\begin{figure}[h!]
\centerline{\scalebox{0.4}{
	\includegraphics{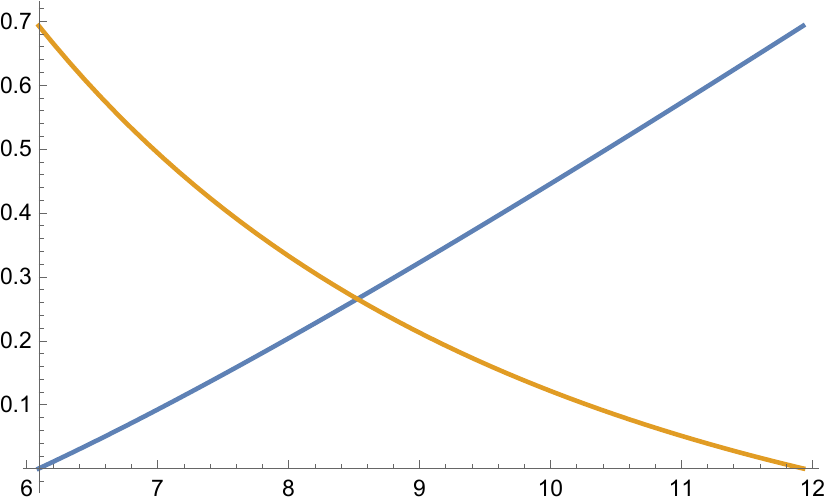}\qquad \qquad \qquad \qquad
    \includegraphics{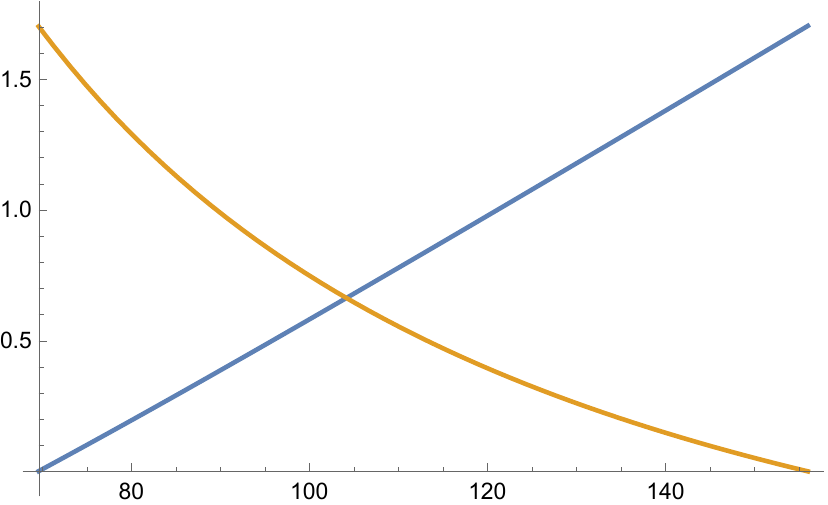}}}
	\caption{Plots of $\cosh{(n\log\rho-\xi_o)}-\alpha$ at the integer vales $n=0$ (blue) and $n=1$ (orange) for $r=1$ and $\kappa=-0.13$ (left), $\kappa=0.06$ (right)  as functions of $c=|\bar{C}_1|$ in the range of allowed values, i.e., $6.1\le c\le 11.9$ (left) and $70\le c\le 156$. The optimal choice for $c$ in each case is at the intersection point, i.e., $\tilde{c}=8.52544$ and $\tilde{c}=104.164$, respectively.}
\label{f:denoms}
\end{figure}

Recalling that $\xi_o=\log \sqrt{\delta_4}$, the inequality \eqref{e:center_cond} can be written as
\begin{equation}
\label{e:center}
\log \rho^{\bar{n}+\Delta}<\log \sqrt{\delta_4}<\log \rho^{\bar{n}+1-\Delta} \quad \Leftrightarrow \quad
\beta\rho^{\bar{n}+\Delta}<|\bar{C}_1|<\beta\rho^{\bar{n}+1-\Delta},
\end{equation}
where we have used the definition of $\sqrt{\delta_4}$ in terms of $\bar{C}_1$ and
\begin{equation}
\beta=\frac{|\bar{C}_1|}{\sqrt{\delta_4}}\equiv \left| \frac{(1+\kappa^2)(1+\kappa r)(\kappa^2-2r\kappa -1)}{\kappa (\kappa -r)(r\kappa^2 +2\kappa -r)}\right|\,.
\end{equation}
Any choice of $c=|\bar{C}_1|$ in the above intervals guarantees that for the corresponding values of $r,\kappa$ the denominator of $Q_n(\tau)$ will not vanish at the integers.
For instance, if one sets $\bar{n}=0$, then $n=0$ and $n=1$ are the integers where the denominator will be the smallest, and one would want to choose $c$ in each range so that $n=0$ and $n=1$ are equidistant from the soliton center. Recall that the soliton center is $\xi_o/\log \rho$, and the choice of $c$ in the above intervals guarantees that
$0<\Delta<\xi_o/\log \rho<1-\Delta<1$. The plots in Fig.~\ref{f:denoms} show $\cosh{(n\log\rho-\xi_o)}-\alpha$ at the integer values $n=0,1$. It is clear that the optimal choice for $c$ in each case is the value at the intersection of the two curves.

\begin{figure}[h!]
\centerline{\scalebox{0.4}{
	\includegraphics{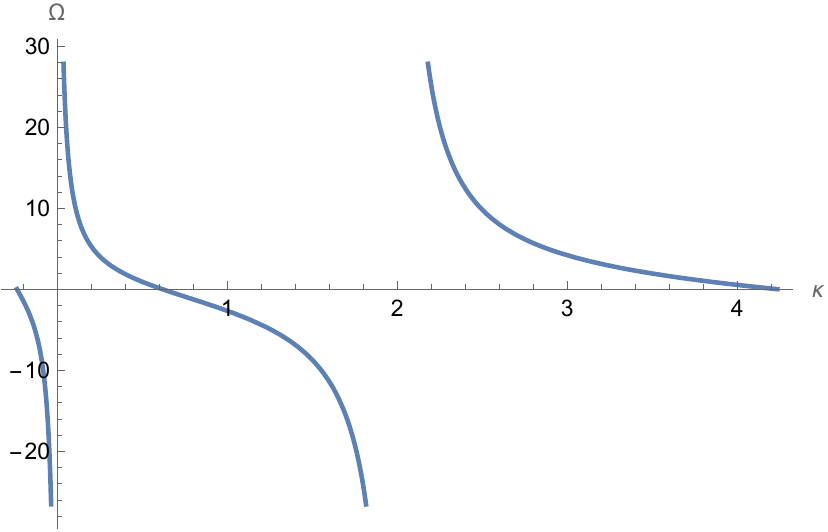}\qquad \qquad \qquad
    \includegraphics{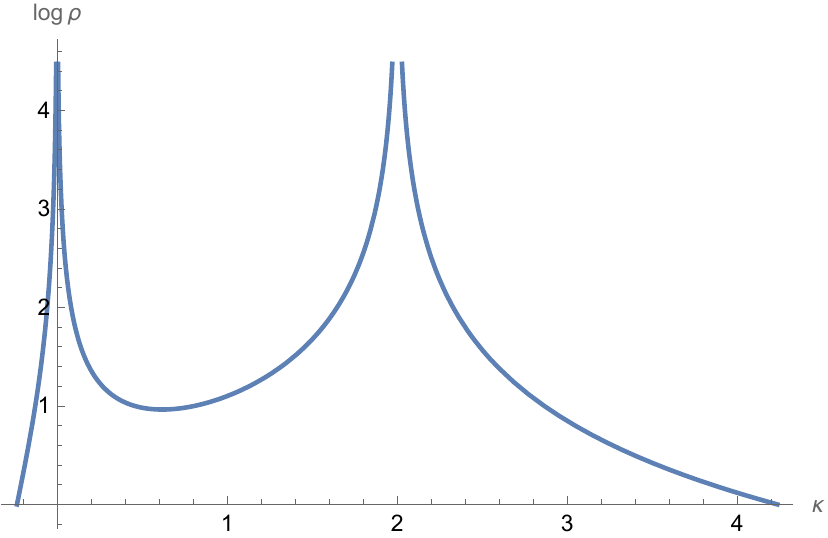}}}
	\caption{Plots of $\bar{\Omega}_1$ (left) and $\log \rho$ (right) as functions of $\kappa \in (-\sqrt{r^2+1}+r,\sqrt{r^2+1}+r)$ for $r=2$.}
\label{f:omega_rho}
\end{figure}

Note that for any fixed $r$ the frequency, $\bar{\Omega}_1$, and $\log \rho$, which controls the width of the breather in each period, become singular at $\kappa=0,r$ (cf Fig~\ref{f:omega_rho}). As it is clear from the contour plots in Fig~\ref{f:contour}, solutions are regular in left and right neighborhoods of $\kappa=0,r$, but in order to obtain solutions whose frequency is not too large and width not too small, one wants to keep sufficiently far from those singular values. On the other hand, the further away one moves from  $\kappa=0,r$ in the intervals where the solution remains bounded for all $n\in \mathbb{Z}$, the closer $\Delta$ is to its limiting value of 1/2, and correspondingly the larger the solution is going to be. So a careful balance is required in the choice of parameters in order to obtain more meaningful KM solutions which are regular for all $n
\in \mathbb{Z}$, $\tau\in \Real$. In Table~1 we provide some such examples, for various values of $r>0$. For the parameters in the Table, we have chosen $\bar{n}=0$ for simplicity.

\begin{table}[h!]
\caption{Soliton parameters for regular KM breathers} 
\centering 
\begin{tabular}{c c c c c c c c c} 
\hline\hline 
$r$ & range for $\kappa$ &$\kappa$ & $\Delta$ & $\bar{\Omega}_1$ & $\log \rho$ & range for $c$ & optimal $c$ & $\inf_{n\in \mathbb{Z}} |\text{denom}| $ \\ [0.5ex] 
\hline 
1/2 & (-0.618, 0)& -0.08 & 0.37 & -3.37 & 3.0 & [90, 193] &  131.7418 & 0.67 \\ 
\ & (0, 0.0665)& 0.05 & 0.44 & 4.53 & 3.8 & [652,1005] & 809.7477 & 0.64 \\
\ &(0.4196, 1/2) & 0.44 & 0.44 & -4.61 & 3.8 & [780,1231] & 980.1981 & 0.67 \\
\ &(1/2, 1.61803) & 0.6  & 0.37 & 3.5  & 3.1 & [130,287] & 193.2707 & 0.71 \\  [.5ex]
\hline 
0.8 & (-0.480625, 0) & -0.1 & 0.32 & -4.0 & 2.3 & [19,41] & 27.6189 & 0.45 \\
\ & (0, 0.09407) & 0.08 & 0.46 & 4.7 & 2.9 & [125,159] & 141.1548 & 0.24 \\
\ & (0.6565247, 0.8) & 0.69 & 0.43 & -5.3 & 3.0 & [188,285] & 231.2327 & 0.4 \\
\ & (0.8, 2.08062) & 1.2  & 0.37 & 1.7  & 1.4 & [9,13] & 10.8274  & 0.12 \\ [.5ex] 
\hline
1 & (-0.414, 0) & -0.13 & 0.31 & -3.5 & 1.8 & [7,11] & 8.5254 & 0.26 \\
\ & (0, 0.10717) & 0.1 & 0.48 & 4.75 & 2.5 & [62,68] & 65.008 & 0.08 \\
\ & (0.806396, 1)  & 0.82 & 0.48 & -4.8 & 2.5 & [102,115] & 108.565 & 0.1 \\
\ & (1, 2.4) & 1.5 & 0.32 & 2.0 & 0.8 & [6.7, 9.9] & 8.14424 & 0.1 \\ [.5ex]
\hline
$\sqrt{5}/2$ & (-0.38, 0)& -0.14 & 0.30 & -3.5 & 1.6 & [3.8,7.1] & 5.18209 & 0.2 \\
\ & (0, 0.113275)  & 0.1 & 0.46 & 5.4 & 2.4 & [45,53] & 48.7821 & 0.13 \\
\ & (0.892, 1.118) & 0.9 & 0.48 & -5 & 2.3 & [80.84] & 82.087 & 0.04 \\
\ & (1.118, 2.61803) & 1.5 & 0.3 & 3.4 & 1.5 & [12,21] & 15.7055 & 0.2 \\ [.5ex]
\hline
2 & (-0.236068, 0)& -0.13 & 0.2 & -5 & 1.0 & [0.7,1.2] & 0.9242 & 0.1 \\
\ & (0, 0.135249) & 0.1 & 0.4 & 10.9 & 1.8 & [10.3, 15.3] & 12.5187 & 0.19 \\
\ & (1.46773, 2) & 1.6 & 0.38 & -11.4 & 1.9 & [37,57] & 45.8334 & 0.2 \\
\ & (2, 4.2) & 3 & 0.2 & 4.2 & 0.8 & [5.1,8.2] & 6.4804 & 0.08 \\
\hline\hline
\end{tabular}
\label{table:nonlin} 
\caption{Parameters for regular KM breathers. Recall that, for fixed $r$, one needs to choose $\kappa\in (-\sqrt{r^2+1}-r,\sqrt{r^2+1}+r)\setminus \left\{0,r\right\}$ in such a way that the regularity condition \eqref{e:Delta} holds. The interval for $c=|\bar{C}_1|$ is then chosen so that \eqref{e:center} holds; for the values in this table, $\bar{n}=0$. The estimate for the lower bound of the denominator is obtained by evaluating $\cosh(n \log \rho-\xi_o)-\alpha $ for the chosen $r,\kappa, c=\tilde{c}$ at $n=0,1$. Finally, choices of $0<\kappa<r$ yield KM2 solutions, while for all other admissible values of $\kappa$ a KM1 solution is obtained.}
\end{table}

\section{Darboux transformations for KM breathers}

A Darboux transformation is meant to take a known solution of an integrable PDE or ODE into a new solution. In \cite{CS2024}, the Darboux transformation for the focusing and defocusing AL equations (AL$_\pm$, respectively) on a nontrivial background were obtained to derive and study the stability of periodic Akhmediev breather solutions. The construction of the Darboux matrix (DM) for the Lax pair \eqref{Lax} follows the one proposed in \cite{G1989,LZ2007} for the case of rapidly decaying solutions (see also \cite{CP2024}, where the Darboux transformation is used to obtain rogue wave solutions of the focusing AL equation superimposed to standing periodic waves).
The DM $D_n(z,\tau)$ takes a fundamental solution $\Psi_n^{[0]}(z,\tau)$ of the Lax pair \eqref{Lax} with potential $Q_n^{[0]}(\tau)$ into a new fundamental solution
$\Psi_n(z,\tau)=D_n(z,\tau)\Psi_n^{[0]}$ corresponding to a potential $Q_n(\tau)$. The requirement that both $\Psi_n^{[0]}$ and $\Psi_n$ satisfy the equations in the Lax pair implies that the DM is
\begin{equation}
\label{e:DE}
D_{n+1}L_n^{[0]}=L_nD_n\,, \qquad \frac{d D_n}{d\tau}=M_n D_n-D_n M_n^{[0]}\,.
\end{equation}
Furthermore, the symmetries of the Lax pair and of the boundary conditions require that
$$
D_n(z,\tau)=\sigma_1 D_n^*(1/z^*,\tau)\sigma_1\,, \qquad \sigma_1=\begin{pmatrix}0 & 1 \\ 1 & 0 \end{pmatrix}.
$$
The details of the construction of the DM are given in \cite{CS2024}; in Appendix~B we report, in our notations, what is needed below in order to reproduce the KM solutions presented in Sec~3, as well as some examples of double KM breathers.

Below, we will apply the DM to the case of KM breathers. In order to compare the solutions in Sec~3 with the ones one can derive using the DM, we need to take into account that the latter is given in terms of the original spectral parameter $z$ (and $\lambda(z)$, as defined in \eqref{lambda}), while the former is expressed in terms of the uniform variable $\zeta$ (cf. Fig~\ref{f:z,zeta}). For this reason, we start here by converting the purely imaginary discrete eigenvalues $\bar{\zeta}_j=i\kappa_j\in D_+$ into the corresponding values for $z_j$ and $\lambda_j$.

\begin{figure}[t]
\centering
	\includegraphics[height=5cm]{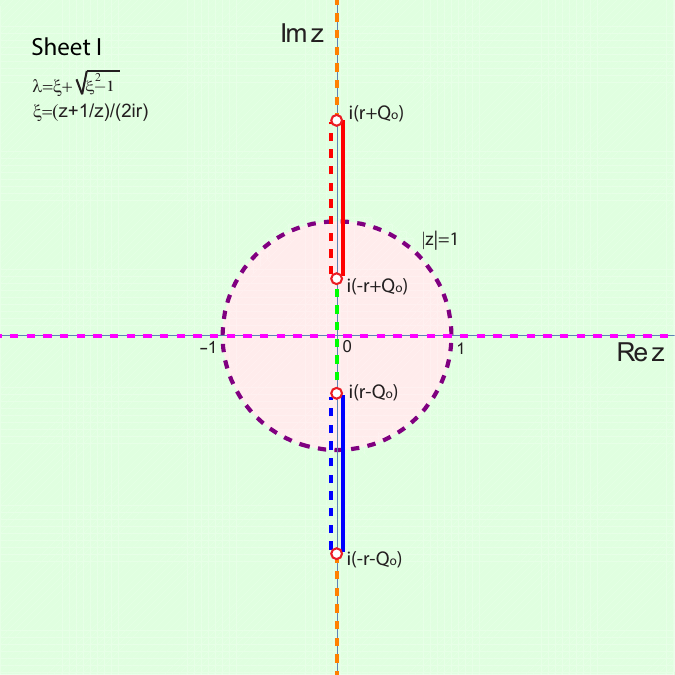}\quad
    \includegraphics[height=5cm]{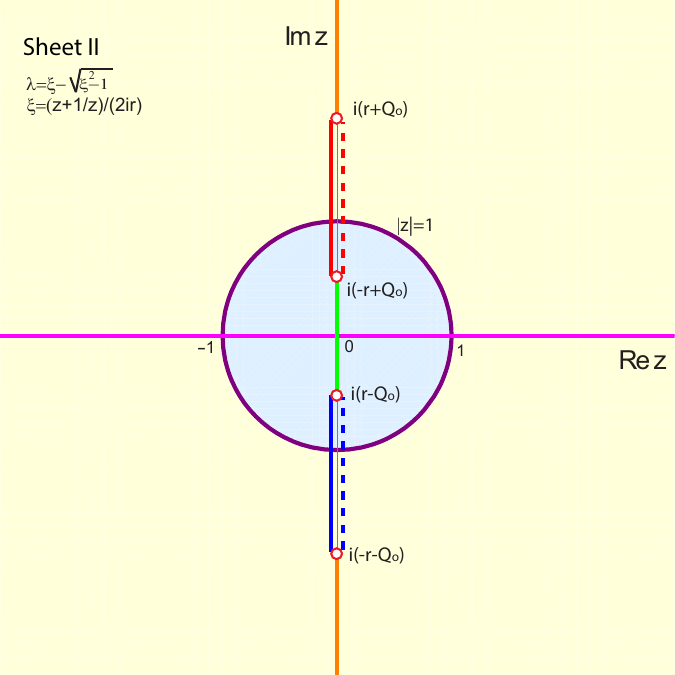} \quad
    \includegraphics[height=5cm]{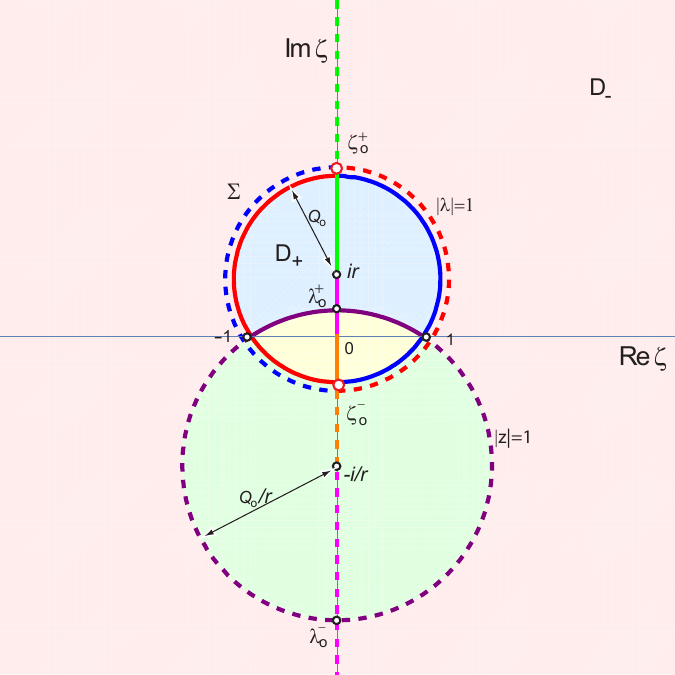}
	\caption{Left: Sheet I of $z$-plane, where $\lambda=\xi+\sqrt{\xi^2-1}$. Center: Sheet II of $z$-plane, where $\lambda=\xi-\sqrt{\xi^2-1}$. Right: $\zeta$-plane; the dashed/solid red and blue circle $|\zeta-ir|=Q_o$ corresponds to the continuous spectrum (where $|\lambda|=1$), and
the corresponding oriented contour $\Sigma$ identifies the regions $D_+$ and $D_-$. The points $\zeta_o^\pm=i(r\pm Q_o)$
are the images of the four branch points $z=i(\pm r\pm Q_o)$, and $\lambda_o^{\pm}=-i(1\mp Q_o)/r$; the dashed/solid purple circle $C=\left\{\zeta\in \Complex:|ir\zeta-1|=Q_o\right\}$
corresponds to $|z|=1$.}
\label{f:z,zeta}
\end{figure}

Consider purely imaginary values of $\zeta=i\kappa$, for all $\kappa$ in the interval:
 $(-Q_o+r, Q_o+r)\setminus\left\{0,r\right\}$. It follows from \eqref{z^2_etc} that
$$
z^2=\frac{r-\kappa}{\kappa(r\kappa+1)}\in \Real\,, \qquad \lambda^2=-\kappa \frac{r-\kappa}{r\kappa +1}\in \Real\,, \qquad z\lambda=i\frac{r-\kappa}{r\kappa+1}\in i\Real\,,
$$
showing that $z^2>0$ in the interval $0<\kappa < r$ where $\lambda^2<0$, whereas  $z^2<0$ for $-Q_o+r < \kappa < 0$ and $r<\kappa < r+\sqrt{1+r^2}$ where $\lambda^2>0$.
This implies that when $z\in \Real$ then $\lambda \in i\Real$ and vice versa. Note also that the third equation above allows one to determine the relative signs of $z$ and $\lambda$.

Specifically, we obtain the following.
\begin{description}
\item{For $0<\kappa < r$:}
\begin{equation}
z=\pm \mu, \qquad \lambda=\pm i \kappa \mu \qquad \text{with }\mu = \sqrt{{\frac{r-\kappa}{\kappa(1+r\kappa)}}}\in \Real^+\,.
\end{equation}
Since $\mu \to 0^+$ as $\kappa\to r^{-}$ and $\mu\to +\infty$ as $\kappa\to 0^+$, considering both signs in $z=\pm \mu$ we see that the entire real $z$-axis corresponds to the segment $i[0,r]$ in the $\zeta$-plane.

\item{For $-Q_o+r<\kappa < 0$:}
\begin{equation}
z=\pm i\mu\,, \qquad \lambda=\pm |\kappa|\mu \qquad \text{with } \mu = \sqrt{\frac{r-\kappa}{|\kappa|(1+r\kappa)}}\in \mathbb{R}^+.
\end{equation}
As $\kappa \to (-Q_o+r)^+$, one has $\mu\to Q_o+r$, and $\mu\to +\infty$ as $\kappa\to 0^-$, so in this case, considering both signs for $z=\pm i\mu$, the segment
$i[-Q_o+r,0]$ in the $\zeta$-plane corresponds to the semilines on the imaginary $z$-axis: $z\in i(-\infty,-(Q_o+r)]\cup i [Q_o+r, +\infty)$.

\item{For $r<\kappa < r+Q_o$:}
\begin{equation}
z=\pm i \mu\,, \qquad  \lambda = \mp \kappa \mu \qquad \text{with } \mu = \sqrt{\frac{|r-\kappa|}{\kappa(1+r\kappa)}}\in \mathbb{R}^+.
\end{equation}
And since  $\mu\to Q_o-r$ as $\kappa \to (r+Q_o)^-$ and $\mu\to 0$ as $\kappa\to r^+$, it follows, again considering both signs in $z=\pm i\mu$ it follows that the segment
$i[r,r+Q_o]$ in the $\zeta$-plane corresponds to segment $i[-Q_o+r,Q_o-r]$ along the imaginary $z$-axis.
\end{description}
For the purpose of the DM, one wants to parametrize both $z$ and $\lambda$ above using one single parameter, and
we can use the same parametrization for $\lambda$ which was used for our KM breathers, namely:
$$
\lambda=\rho^{-1/2}e^{i\phi/2}\equiv e^{-\log \sqrt{\rho}+i\phi/2}\,, \qquad \lambda^{\pm n}e^{\pm i\Omega \tau}=\exp{\left[\mp n\log\sqrt{\rho}\mp (\Im \Omega) \tau \pm i \frac{n \phi}{2}\pm i(\Re \Omega)\tau \right]},
$$
and further simplify the latter to:
$$
\lambda^{\pm n}e^{\pm i\Omega \tau}=\exp{\left[\mp n\log\sqrt{\rho} \pm i \frac{n \phi}{2}\pm i \Omega \tau \right]}
$$
based on the fact that for KM breathers $\rho>1$, $\phi/2=0,\pi, \pm \pi/2$ depending on which of the above cases for $\kappa$ we are considering, and also that $\Im \Omega =0$. As far as $z$ is concerned, that can be obtained from \eqref{lambda}, and the appropriate sign chosen according to the cases indicated before.
We need to establish the range of values of $\rho>1$. This can be done by looking at each interval of values for $\kappa$ below, determine what the corresponding interval of values for $\rho$ should be by solving \eqref{e:rhok} for $\rho$ in terms of $\kappa$.
For any fixed $r$, computing $\partial_\kappa \rho$ when $0<\kappa<r$ yields two stationary points, located at $\kappa =(\pm Q_o-1)/r$. These correspond to $\zeta=i\kappa $ coinciding with the points $\lambda_o^\pm$ (see Fig.~\ref{f:z,zeta}); and $\kappa=(Q_o-1)/r\in (0,r)$. In conclusion, for each given $r$, one has:
\begin{equation}
\label{e:minrho}
\min_{0<\kappa<r}\rho(r,\kappa)=\rho(r,(Q_o-1)/r)=\frac{r^2}{(Q_o-1)^2}>1.
\end{equation}
On the other hand, when $-Q_o+r<\kappa<0$ or $r<\kappa<Q_o+r$, one finds that $\partial_\kappa \rho=(-r^2\kappa^2-2\kappa+r)/\kappa^2(\kappa-r)^2$, with the same two stationary points $\kappa=-i \lambda_o^\pm$. In these cases, however, the stationary points are both outside the ranges of values for $\kappa$, so $\rho$ is a monotonic function of $\kappa$. Furthermore, clearly $\rho \to +\infty$ as either $\kappa\to 0^-$ or as $\kappa\to r^-$, while $\rho\to 1$ as $\kappa\to r\pm Q_o$, confirming that in both cases $\rho$ can take any value greater than 1.

\begin{description}
\item{$0<\kappa<r$:}

The KM breathers in this case have $\lambda$ purely imaginary and $z$ real, and $z, -i\lambda$ with the same sign. Based on the fact that we expect $|\lambda|<1$ and taking,
without loss of generality, $\lambda$ in the upper half-plane, we have
\begin{equation}
\label{e:zKM2}
\lambda=i/\sqrt{\rho}\equiv i e^{-\log \sqrt{\rho}}
\quad \Leftrightarrow \quad z=r\sinh \log\sqrt{\rho} \pm \sqrt{r^2\sinh^2\log \sqrt{\rho}-1}.
\end{equation}
Now, it is not obvious that $z$ above is indeed real for any choice of $\rho>1$; the condition that the argument of the square root be non-negative can be written as
$r^2(\rho+1/\rho+2)-4Q_o^2\ge 0$, which, in turn is equivalent to
$$
r^2\rho^2-2(1+Q_o^2)\rho+r^2\ge0.
$$
The roots of the above quadratic equation are $\rho=(1\pm Q_o)^2/r^2$, and the inequality is satisfied if either $\rho\le (Q_o-1)^2/r^2$ or $\rho\ge (Q_o+1)^2/r^2$
and the latter can be shown to coincide with the minimum value of $\rho$ in the interval $0<\kappa<r$, as computed in \eqref{e:minrho}, so indeed $z$ as given in \eqref{e:zKM2} is real. Note that both choices for the sign of the square root in \eqref{e:zKM2} would correspond to positive $z$ (equivalently, to positive $-i\lambda z$), and therefore for a given $\rho$ greater or equal than the minimum value in \eqref{e:minrho}, one could have two different KM breathers.

\item{$-Q_o+r<\kappa<0$:}

The KM breathers in this case have $\lambda$ real and $z$ purely imaginary, and, as before, $\lambda, -iz$ have the same sign. Using similar parametrization as before:
\begin{equation}
\label{e:zKM1n}
\lambda=1/\sqrt{\rho}\equiv e^{-\log \sqrt{\rho}}
\quad \Leftrightarrow \quad z=ir\cosh \log\sqrt{\rho} + i\sqrt{r^2\cosh^2\log\sqrt{\rho}+1}
\end{equation}
and here the positive sign of the square root is the only one that ensures the correct sign of $z$.
\item{$r<\kappa<Q_o+r$:}

The only difference with respect to the previous case is that $-iz$ and $\lambda$ should have opposite signs. If we want to keep $z$ in the upper half plane, then we should have
\begin{equation}
\label{e:zKM1p}
\lambda=-e^{-\log \sqrt{\rho}}
\quad \Leftrightarrow \quad z=-ir\cosh \log\sqrt{\rho} +i\sqrt{r^2\cosh^2\log\sqrt{\rho}+1}
\end{equation}
\end{description}
In all cases $\Omega$, as a function of $z,\lambda$ is given in Eq.~\eqref{e:Omega}, which is actually independent of the signs since it depends on $z^2$ and $\lambda^2$.

In order to build the single KM and double KM
breathers using the DM, we take $Q_n^{[0]}(\tau)=Q_o$, i.e, the background solution. In this case, one can take the fundamental matrix solution of the Lax pair to be
\begin{equation}
\Psi_n^{[0]}(\rho,\tau)=(ir)^n\begin{pmatrix} \lambda(\rho)^n e^{i\Omega(\rho)\tau} & \frac{Q_o}{ir/\lambda(\rho)-z(\rho)}  \lambda(\rho)^{-n} e^{-i\Omega(\rho)\tau}  \\
\frac{Q_o}{ir\lambda(\rho)-1/z(\rho)} \lambda(\rho)^n e^{i\Omega(\rho)\tau} &  \lambda(\rho)^{-n} e^{-i\Omega(\rho)\tau}
\end{pmatrix},
\end{equation}
where $\lambda(\rho)$ and $z(\rho)$ are given by \eqref{e:zKM2}, \eqref{e:zKM1n}, or \eqref{e:zKM1p} depending on the desired range of values for $\kappa$; and $\Omega(\rho)$ is defined by \eqref{e:Omega}.

Following the construction of the Darboux transformation outlined in Appendix A, after simplification one can obtain explicit single KM solutions like those given in Section 3. These will depend on a single real parameter $\rho_{1}$ that determines $z_{1}\equiv z(\rho_{1})$, $\lambda_{1}\equiv\lambda(\rho_{1})$, and $\Omega_{1}\equiv\Omega(\rho_{1})$; as well as a complex proportionality constant $\gamma_{1}\equiv|\gamma_{1}|e^{i\varphi}$ (defined in \eqref{e:DMprop}) which plays the role of the norming constant.

First, by parametrizing $z(\rho)$ and $\lambda(\rho)$ through either \eqref{e:zKM1n} or \eqref{e:zKM1p} (i.e. enforcing that $z_{1}=-z_{1}^{*}$ and $\lambda_{1}=\lambda_{1}^{*}$), one can obtain the solution
\begin{equation}
\label{darboux_solution_1}
Q_{n}(\tau)=Q_{o}+\frac{r}{2}(\lambda_{1}-\lambda_{1}^{-1})B\frac{(A-A^{-1})\cos(\eta-\varphi)-i(A+A^{-1})\sin(\eta-\varphi)}{\cosh(\xi-\xi_{o})+B\sin(\eta-\varphi)}\,,
\end{equation}
where $\xi=n\log\rho_{1}$, $\eta=2\Omega_{1}\tau$, and
\begin{equation}
    A=\frac{Q_{o}}{ir\lambda_{1}-z_{1}^{-1}}\,,\;\;B=\frac{(z_{1}+z_{1}^{-1})A}{|z_{1}-A^{2}z_{1}^{-1}|\sqrt{\rho_{1}}}\,,\;\;\xi_{o}=-\log|\gamma_{1}|\sqrt{\rho_{1}}\,.
\end{equation}
If instead $z(\rho)$ and $\lambda(\rho)$ are parametrized by \eqref{e:zKM2} with either choice of the sign of the square root (i.e. $z_{1}=z_{1}^{*}$ and $\lambda_{1}=-\lambda_{1}^{*}$), the solution is the KM2 breather, in the form
\begin{equation}
\label{darboux_solution_2}
Q_{n}(\tau)=Q_{o}+(-1)^{n}\frac{r}{2}(\lambda_{1}-\lambda_{1}^{-1})B\frac{(A-A^{-1})\cos(\eta-\varphi)-i(A+A^{-1})\sin(\eta-\varphi)}{\cosh(\xi-\xi_{o})+(-1)^{n}B\sin(\eta-\varphi)}\,.
\end{equation}

For the purpose of comparing the solutions \eqref{darboux_solution_1} and \eqref{darboux_solution_2} with the previous forms \eqref{e:KM1} and \eqref{e:KM2}, one can set the phase of $\gamma_{1}$ to be $\varphi=\pi/2$ (which is apparently equivalent to setting the phase of $\bar{C}_{1}$ to zero as was done in Section 3). With this, \eqref{darboux_solution_1} can be rewritten as
\begin{equation}
\label{darboux_solution_3}
Q_{n}(\tau)=Q_{o}+\frac{ir}{2}(\lambda_{1}-\lambda_{1}^{-1})B\frac{(A+A^{-1})\cos\eta-i(A-A^{-1})\sin\eta}{\cosh(\xi-\xi_{o})-B\cos\eta}\,,
\end{equation}
from which by comparison with \eqref{e:KM1} one can directly identify the coefficients in terms of the uniformization variable (parametrized by $\kappa$) with the parameters in terms of the original spectral parameter (parametrized by $\rho_{1}$) in the following way:
\begin{equation}
    (\lambda_{1}-\lambda_{1}^{-1})B(A\pm A^{-1})\longleftrightarrow\frac{\delta_{1}^{*}\pm\delta_{2}^{*}}{\sqrt{\delta_{4}}}\,,\quad \quad B\longleftrightarrow\alpha\,.
\end{equation}
Finally, by comparing the two expressions for the center of the KM breather $\xi_{o}$, one finds a way to relate the norming constant $\bar{C}_{1}$ to the proportionality constant $\gamma_{1}$; namely, $|\gamma_{1}|^{2}\rho_{1}=1/\delta_{4}$ (recall that $\delta_{4}$ is proportional to $|\bar{C}_{1}|^{2}$).

The parameter $B$ as defined above is real and could be positive or negative depending on the choice of parametrization. Regardless, it is true that $|B|>1$ for any $\rho_{1}$, in which case the solutions above are generically singular. Following the same logic as in Section 4, we seek  parameters $\rho_{1}$ and $\gamma_{1}$ such that no integer lies in the interval where $\cosh(\xi-\xi_{o})\leq|B|$. A necessary condition for regularity that ensures that the width of this interval is less than 1 is
\begin{equation}
\label{darboux_width_condition}
    \arccosh|B|-\frac{1}{2}\log\rho_{1}<0\,.
\end{equation}
As before, it may be deemed to be useful to impose a stronger regularity condition, for example by replacing the $1/2$ in \eqref{darboux_width_condition} by $1/4$~\footnote{However, as
regards the practical manifestation of the instability, as illustrated
in the next section, this was not found to modify the observed
dynamics significantly.}. Furthermore, we can center the solution between two lattice sites $\bar{n}$ and $\bar{n}+1$ by choosing $|\gamma_{1}|$ in the interval
\begin{equation}
\label{darboux_center_condition}
    \rho_{1}^{\Delta-\bar{n}-\frac{3}{2}}<|\gamma_{1}|<\rho_{1}^{-\Delta-\bar{n}-\frac{1}{2}}\,,\;\;\Delta=\frac{\arccosh|B|}{\log\rho_{1}}\,.
\end{equation}

\begin{figure}[h!]
\includegraphics[width=\textwidth]{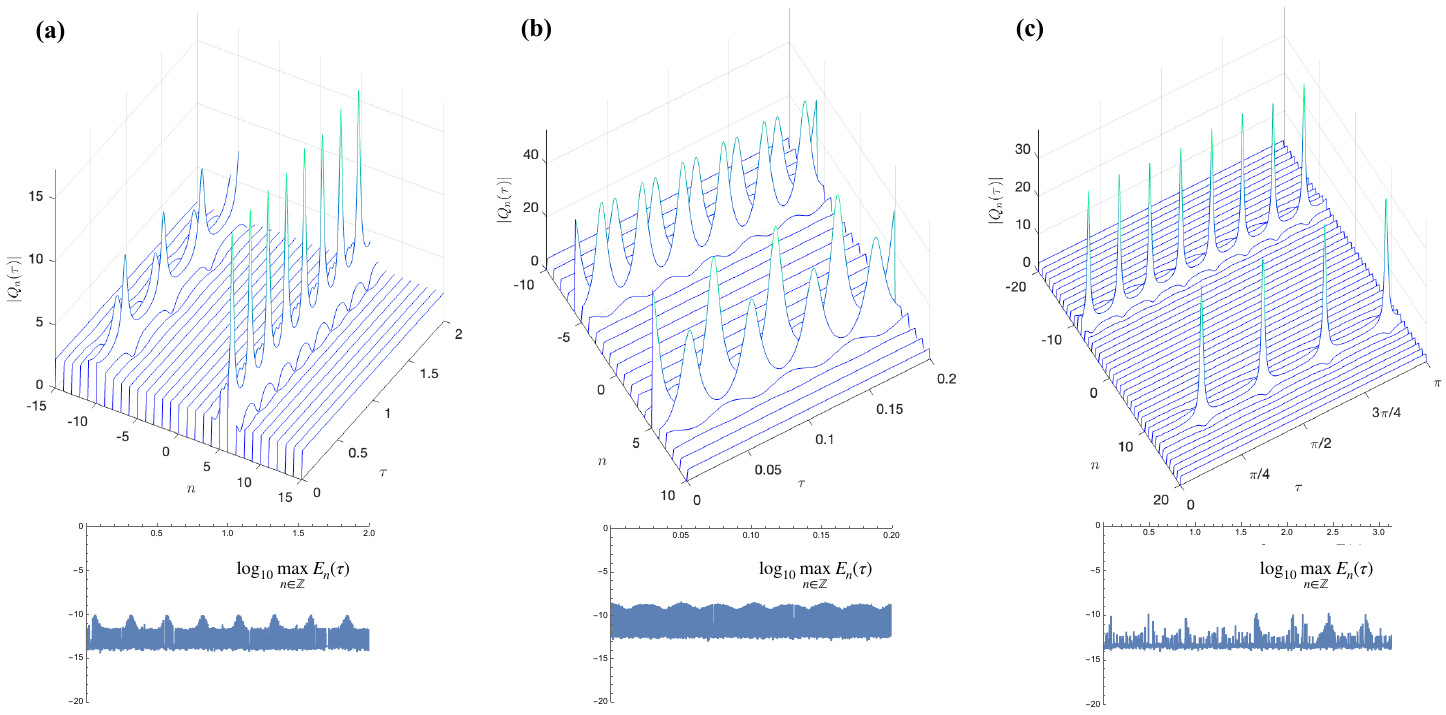}
\centering
\caption{\textbf{(a)} Double KM1 solution with $r=2$ and $z_{1}$ and $z_{2}$ parametrized by \eqref{e:zKM1n}, corresponding to $\rho_{1}=6$, $\gamma_{1}=10^{-4}$, $\rho_{2}=3$, $\gamma_{2}=10^{4}$. \textbf{(b)} Double KM2 solution with $r=4$ and $z_{1}$ and $z_{2}$ parametrized by \eqref{e:zKM2} with the positive sign, corresponding to $\rho_{1}=8$, $\gamma_{1}=10^{-4}$, $\rho_{2}=12$, $\gamma_{2}=10^{4}$. \textbf{(c)} Double KM1 solution with $r=2$ and $z_{1}$ and $z_{2}$ parametrized by \eqref{e:zKM1n}, corresponding to $\rho_{1}=2.2588611606273092$, $\gamma_{1}=10^{-4}$, $\rho_{2}=3.9647690662942265
$, $\gamma_{2}=10^{4}$; chosen such that the period of the left wave is $\pi/8$ and the period of the right wave is $\pi/4$.
}
\label{2KM_Figure_extra}
\end{figure}

Additionally, the Darboux transformation can be used to construct a double KM solution according to the formulae provided in Appendix A. We remark that if two parameter sets $\{\rho_{1},\gamma_{1}\}$ and $\{\rho_{2}, \gamma_{2}\}$ are chosen according to the regularity conditions \eqref{darboux_width_condition} and \eqref{darboux_center_condition} such that each set separately characterizes a well-behaved single KM solution, the resulting double KM solution is not guaranteed to remain regular (even with centers $\bar{n}_{1}$ and $\bar{n}_{2}$ chosen to be sufficiently separated). However, one can find combinations of parameters for which the double KM solution is well-behaved, examples of which are displayed in Fig.~\ref{2KM_Figure_extra}.

\section{Instability \& numerical blow-up}

In this section, we study numerically and show respective results for a few representative cases of regular single KM breathers associated with Fig.~\ref{1KM_Figure}.
Specifically, we attempt herein to shed light on the spectral stability of KM breathers that are
introduced in this work.~In doing so, we obtain first the respective time-periodic yet numerically exact
solution of Eq.~\eqref{e:AL} by using Newton's method which is subsequently used
in the variational equations
towards deriving the monodromy matrix.~The eigenvalues of the monodromy matrix are the Floquet
multipliers which provide information about the stability of the pertinent structures, see~\cite{SCCKK2020}
for the formulation of the stability analysis problem.~Briefly, and as far as Newton's method is concerned,
time-periodic solutions of period $T_{b}=\pi/\tilde{\Omega}_{1}$
\begin{align}
Q_{n}=\sum_{m=-k_{m}}^{k_{m}}{\cal Q}_{n,m}e^{2\mathrm{i}m\tilde{\Omega}_{1}\tau},
\label{eq:fourier_decomp}
\end{align}
where ${\cal Q}_{n,m}$ are the Fourier coefficients, and $2k_{m}+1$ are the total Fourier
modes in time we consider.~In this work, we use $k_{m}=61$ and thus $123$ Fourier collocation
points are used for the temporal discretization.~Upon plugging Eq.~\eqref{eq:fourier_decomp} into Eq.~\eqref{e:AL}
where periodic (in space) boundary conditions are imposed therein, we obtain a
nonlinear algebraic system of equations for ${\cal Q}_{n,m}$ that we solve with
Newton's method; see~\cite{SCCKK2020} for the formulation of the relevant root-finding
problem and~\cite{CJMHKK2024} for a recent yet relevant application of the method.
We mention in passing that the initial guesses to the solver are provided directly by the
exact solutions we aim to study, and Newton's method converges in 1 iteration (within machine precision).
However, we have noticed that the solver may need to perform more iterations, 
e.g., 3-4, to converge for the same number of Fourier collocation points employed.~One such case
is strongly reminiscent of Fig.~\ref{1KM_Figure}(c) but with $r=2$, $\kappa=3$,
and $c=6.48041$ where the solver converged in 3 iterations (if $k_{m}=81$, i.e., $163$
collocation points in time, the solver convergences in 1 iteration).~There are also
other cases, such as the one with $r = 4$, $\kappa = 8$, and $c=0.253$ where our
nonlinear solver failed to converge entirely.~This observation raises an important
problem that merits a further yet separate investigation in developing suitable computational
methods for identifying such periodic orbits. 
Regardless, and for the single KM solutions we present herein, we found
that the use of $123$ collocation points in time together with a lattice of $100$ sites accurately resolves the time-translation
mode $1+0\mathrm{i}$ within $10^{-7}-10^{-8}$ error.

Finally, we provide relevant dynamics of the
KM solutions that are numerically obtained in order
to explore the corresponding evolution of such states.
For these direct numerical computations,
we use the Burlisch-Stoer method~\cite{HNW2008} with $10^{-14}$ relative and absolute errors for all time
integrations (including those that are needed for the Floquet computations, i.e., variational problem).
As a diagnostic in order to assess the accuracy of the
dynamical results, we leverage the integrable structure
of the AL model and, in particular, its bearing of an infinite
number of associated conservation laws.
The conserved quantities in the case of waveforms whose
tails asymptotically vanish can be derived recursively
using formulas (3.4.155) and (3.4.157) in \cite{APT2004}, and simplified using the summation by parts formula (A.1.2).
The conserved quantities for non-zero boundary conditions can be obtained by subtracting the boundary conditions to ensure all the series are convergent.
Specifically, we have
\begin{gather}
\Gamma_1=\sum_{n=-\infty}^{+\infty}\left\{ Q_n Q_{n-1}^*-Q_o^2 \right\}\,,
\qquad \bar{\Gamma}_1=\sum_{n=-\infty}^{+\infty}\left\{ Q_{n-1}Q_{n}^*-Q_o^2 \right\},
\label{eq:consv_Gamma_1}
\end{gather}
\begin{gather}
\Gamma_2=\sum_{n=-\infty}^{+\infty}\left[ Q_nQ_{n-2}^*\left(1-|Q_{n-1}|^2 \right)-\frac{1}{2}(Q_{n}Q_{n-1}^{*})^2 -Q_o^2+\frac{3}{2}Q_o^4\right], \\
\bar{\Gamma}_2=\sum_{n=-\infty}^{+\infty}\left[ Q_{n-2}Q_{n}^*\left(1- |Q_{n-1}|^2 \right)-\frac{1}{2}(Q_{n-1}Q_{n}^{*})^{2}-Q_o^2+\frac{3}{2}Q_o^4\right].
\label{eq:consv_Gamma_2}
\end{gather}
The IST also shows the following infinite product is conserved:
\begin{gather}
c_{\infty}=\prod_{n=-\infty}^{+\infty}\frac{|Q_n|^2-1}{Q_o^2-1}.
\label{eq:consv_cinf}
\end{gather}
Finally, from the expression of the Hamiltonian in the case of zero boundary conditions, once the nontrivial boundary conditions are subtracted out we obtain (cf. also \cite{VK1992}):
\begin{gather}
\label{e:H}
H=\sum_{n=-\infty}^{+\infty}\left\{ Q_n^* \left(Q_{n+1}+Q_{n-1}\right)-2Q_o^2\right\}+2\sum_{n=-\infty}^{+\infty}\log \frac{|Q_n|^2-1}{Q_{o}^{2}-1}.
\end{gather}
with
\begin{equation}
    H=\Gamma_{1}+\bar{\Gamma}_{1}+2\log c_{\infty}.
    \label{eq:consv_Ham}
\end{equation}
It is worth pointing out that both the infinite product \eqref{eq:consv_cinf} and the related log series in \eqref{e:H} satisfy the necessary condition for convergence, since $|Q_n|\to Q_o$ as $n\to \pm \infty$ for both $Q_o<1$ and $Q_o>1$. In the case of zero boundary conditions, the defocusing AL lattice
conserves the functional $\mathcal{P} =\sum_{n=-\infty}^{\infty}\log (1-|Q_n|^2)$, and global existence is ensured if the initial condition
satisfies $||Q_n(0)||_\infty < 1$, but to the best of our knowledge the analog of this result has not yet been established in the case of nonzero boundary conditions even when $0<Q_o<1$.
For our exact solutions, we observe that when $Q_o>1$,
even though $|Q_n|$ can dip below $Q_o$ at some lattice sites,
one has $|Q_n|>1$ for all $n\in \mathbb{Z}$. In this case, the terms in the infinite product \eqref{eq:consv_cinf} and the arguments of the logarithm in \eqref{e:H}  are positive, and requiring
$Q_n-Q_o \in \ell^1$ is sufficient to guarantee the convergence of both \eqref{eq:consv_cinf} and \eqref{e:H}.
It is interesting to add here that we have
observed that in the course of the numerical evolution
dynamics of such unstable solutions, round-off error
may lead, through the resulting destabilization,
to the indefinite growth of the modulus
of particular nodes. As this happens, we have noticed that
(pairwise) some sites drop below $Q_o$ and indeed approach unit modulus, in a way that can be seen
to conserve both the infinite product of \eqref{eq:consv_cinf} and the energy of \eqref{e:H}. How this collapse dynamics manifests itself and its bearing on the AL lattice conservation laws are clearly topics meriting further investigation.

\begin{figure}[ht]
\centering
	\includegraphics[width=0.23\textwidth]{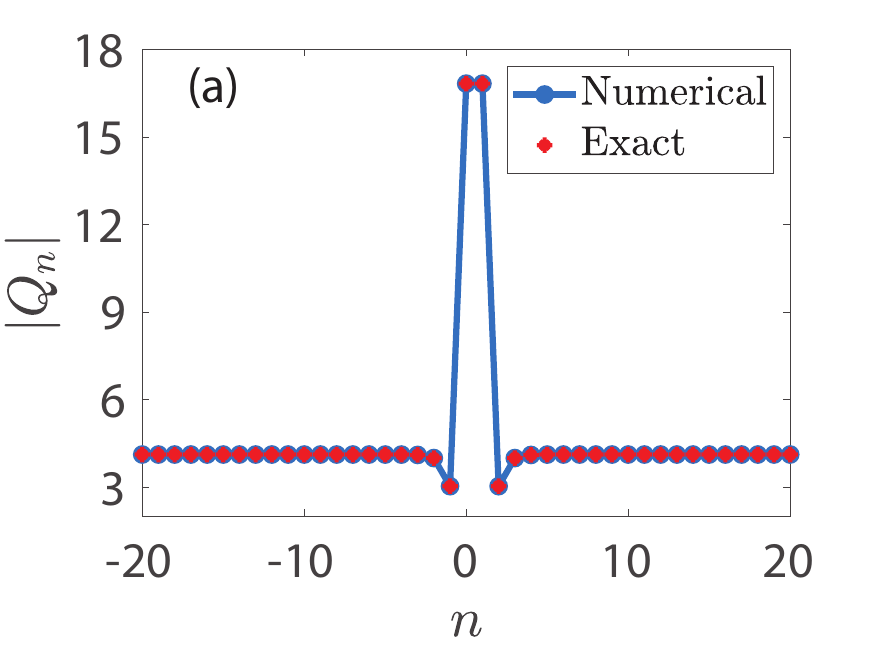}\quad
    \includegraphics[width=0.23\textwidth]{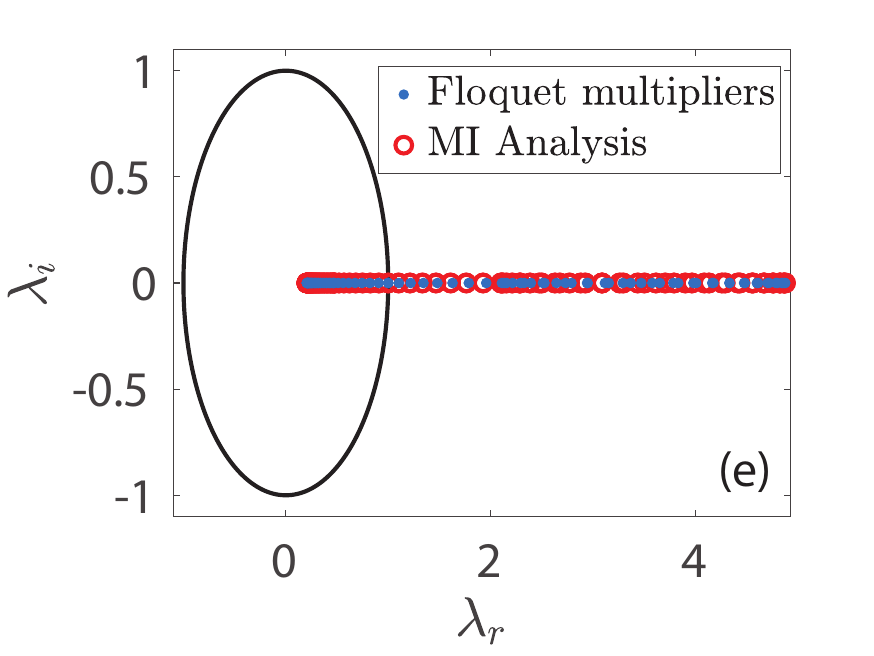} \quad
    \includegraphics[width=0.23\textwidth]{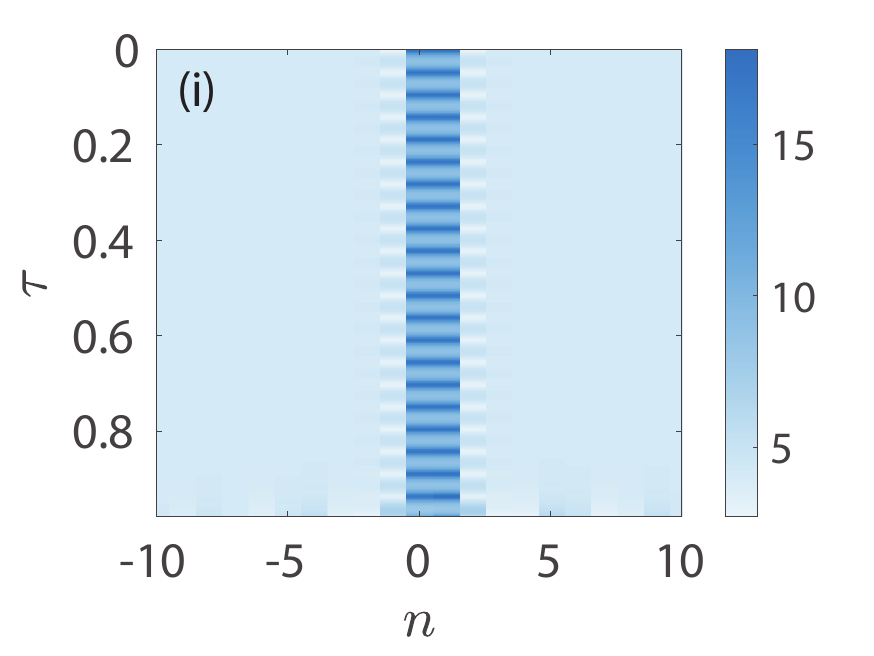}\quad
    \includegraphics[width=0.23\textwidth]{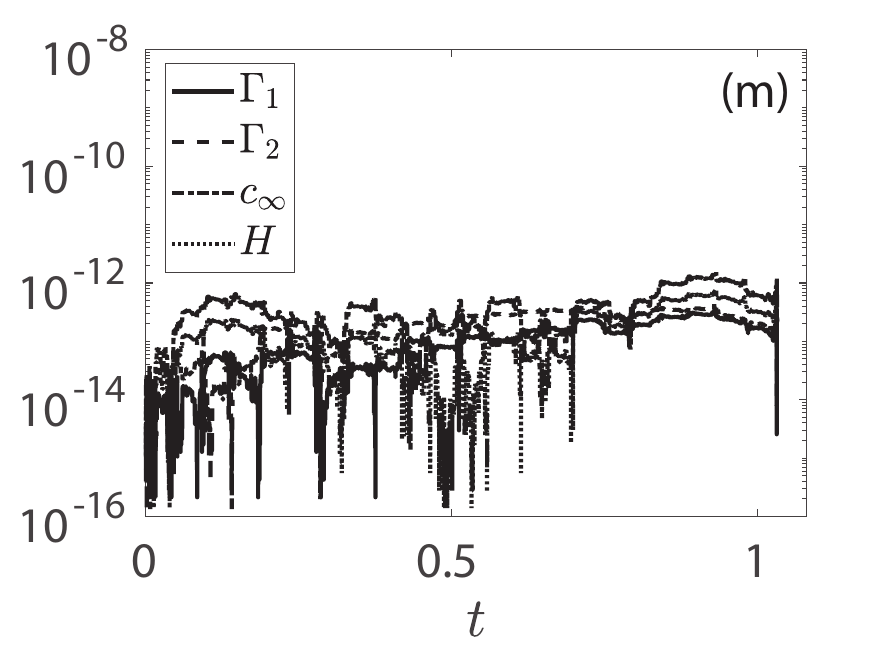}\quad \\
    \includegraphics[width=0.23\textwidth]{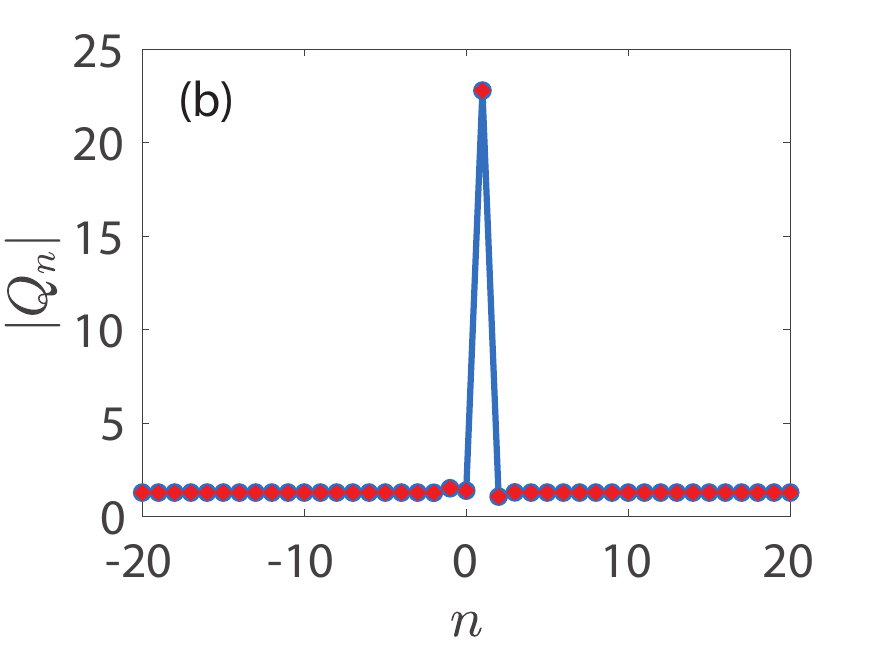}\quad
    \includegraphics[width=0.23\textwidth]{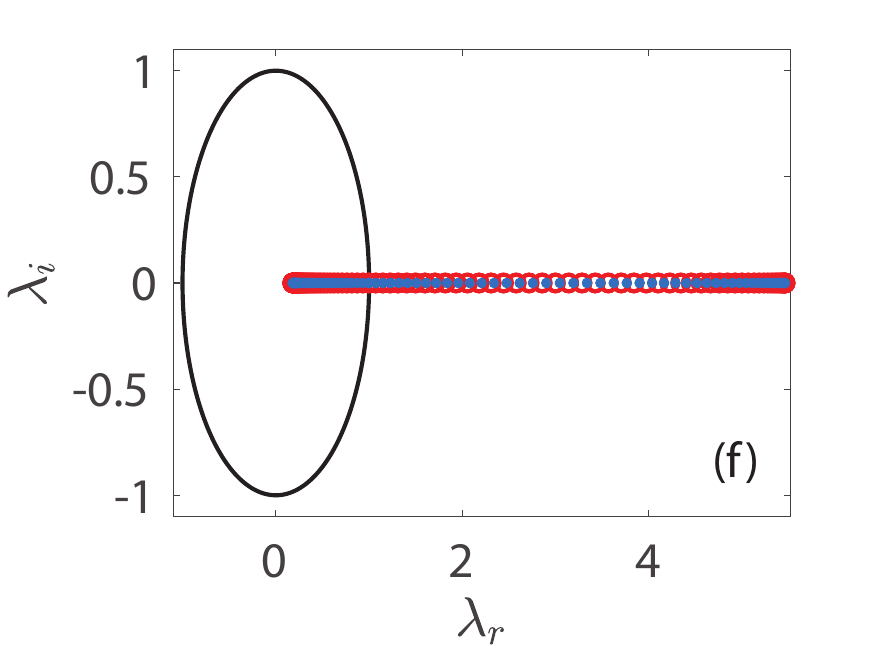} \quad
    \includegraphics[width=0.23\textwidth]{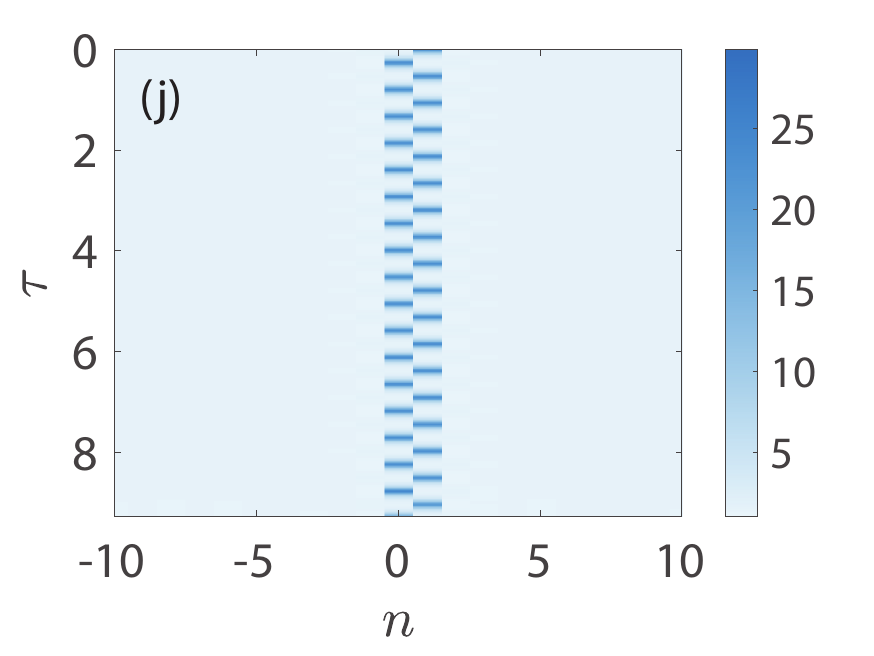}\quad
    \includegraphics[width=0.23\textwidth]{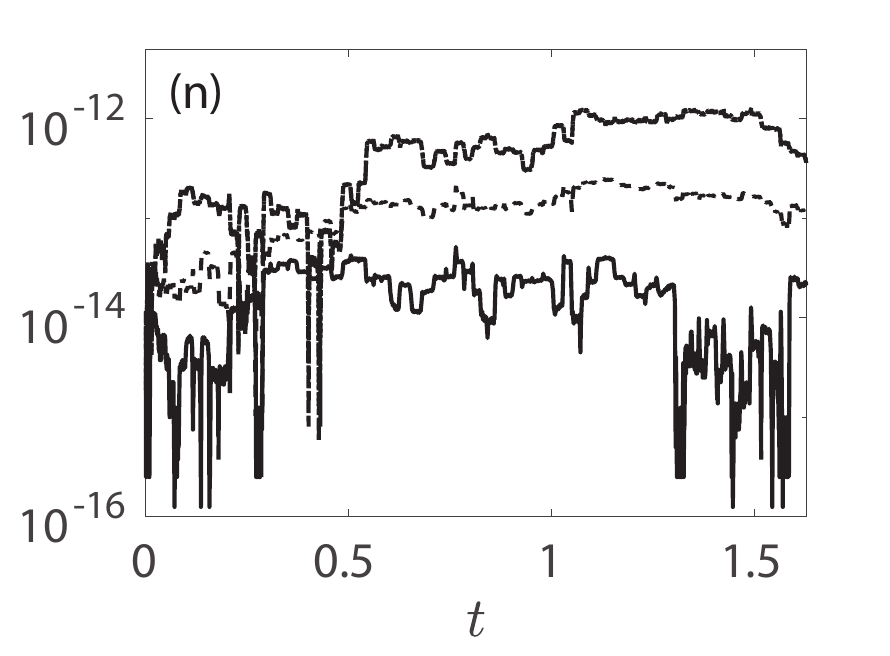}\quad \\
    \includegraphics[width=0.23\textwidth]{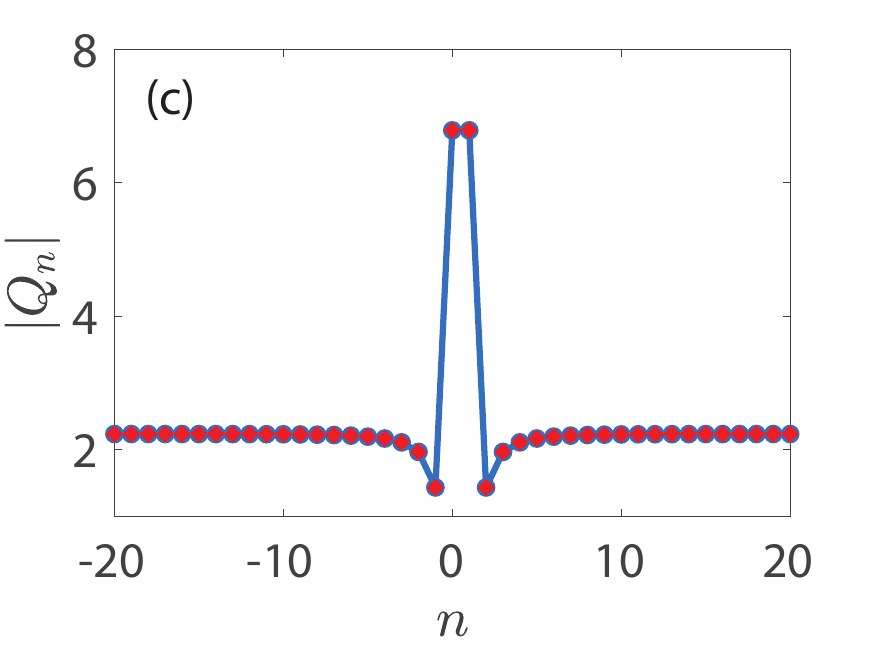}\quad
    \includegraphics[width=0.23\textwidth]{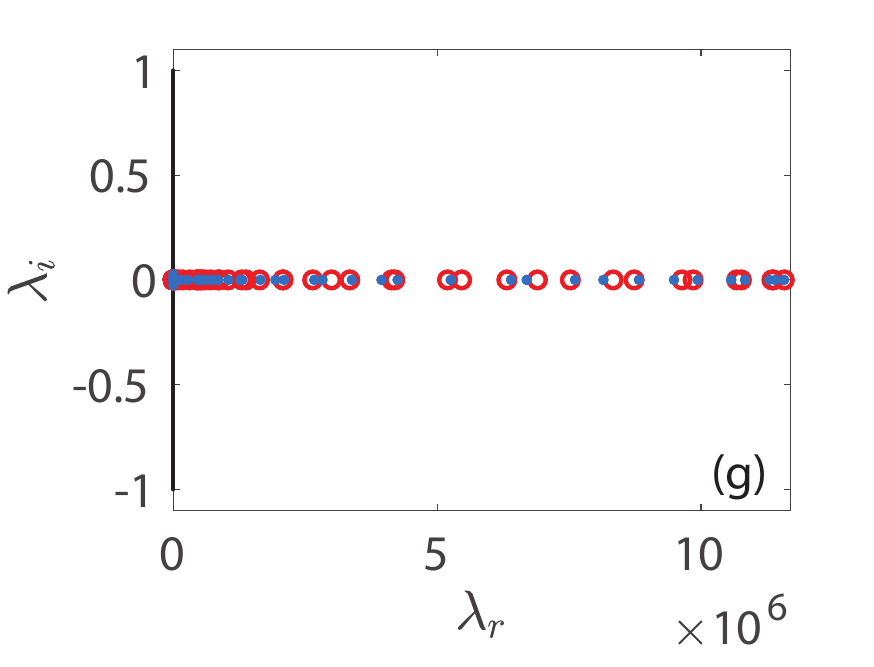} \quad
    \includegraphics[width=0.23\textwidth]{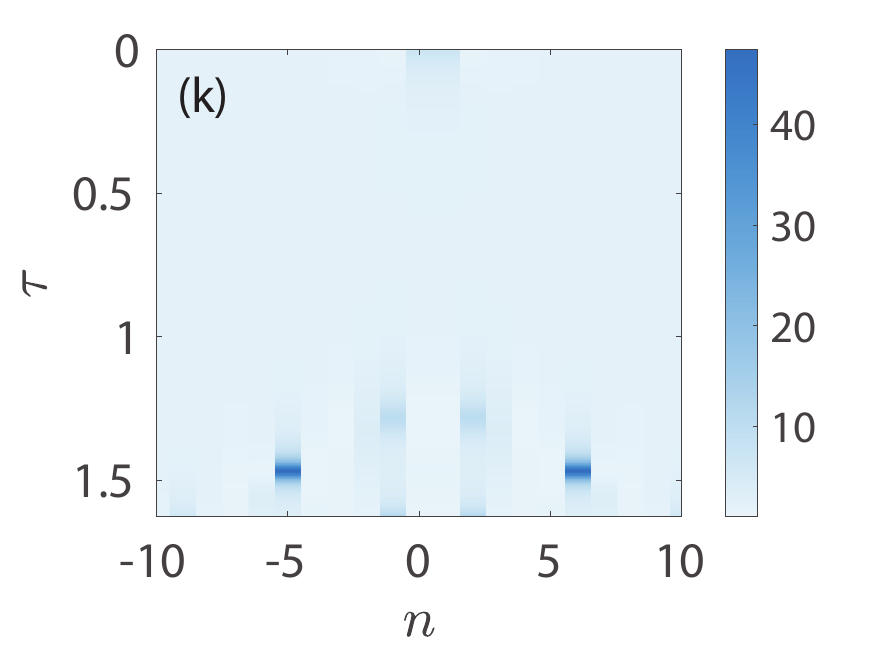}\quad
    \includegraphics[width=0.23\textwidth]{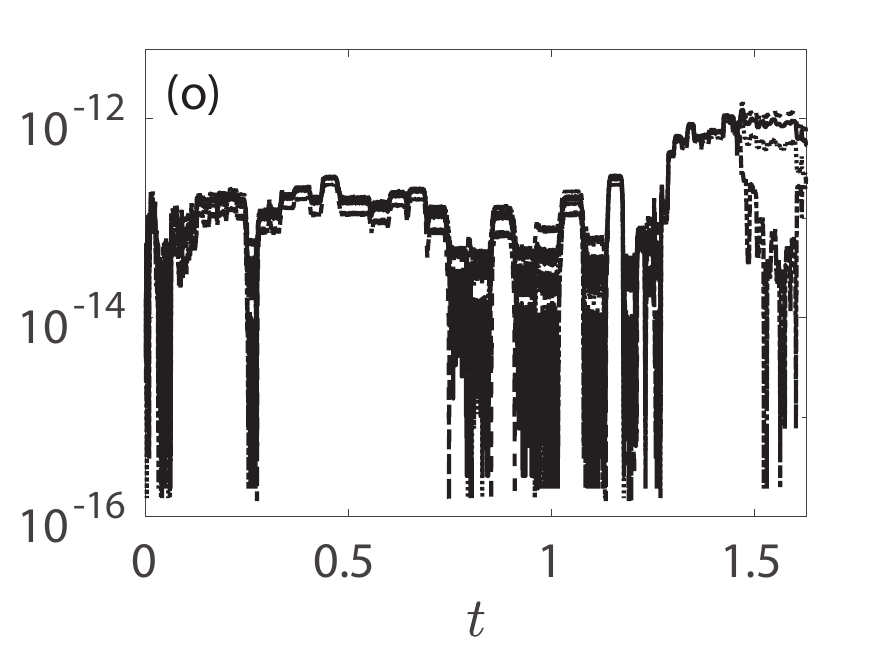}\quad \\
    \includegraphics[width=0.23\textwidth]{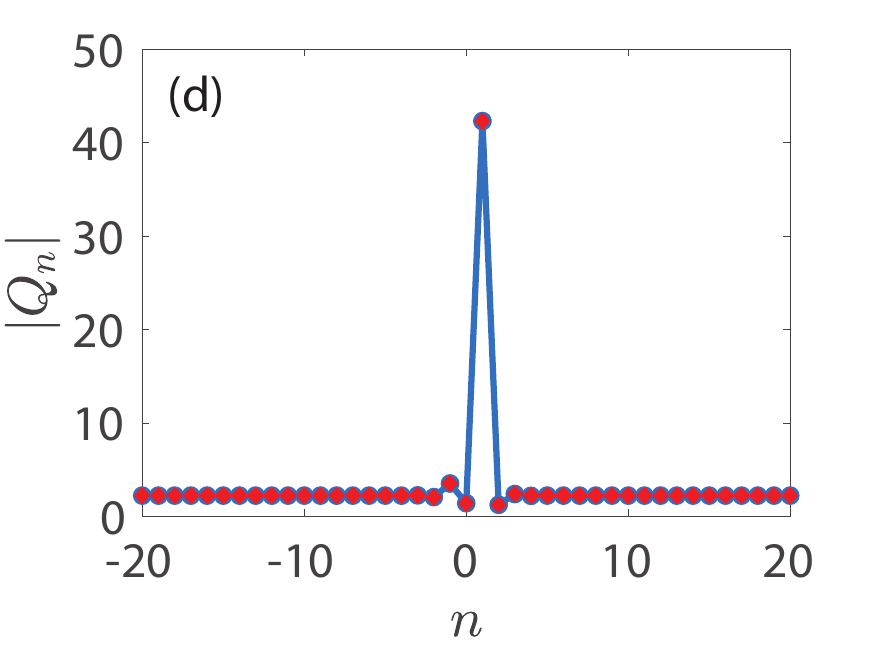}\quad
    \includegraphics[width=0.23\textwidth]{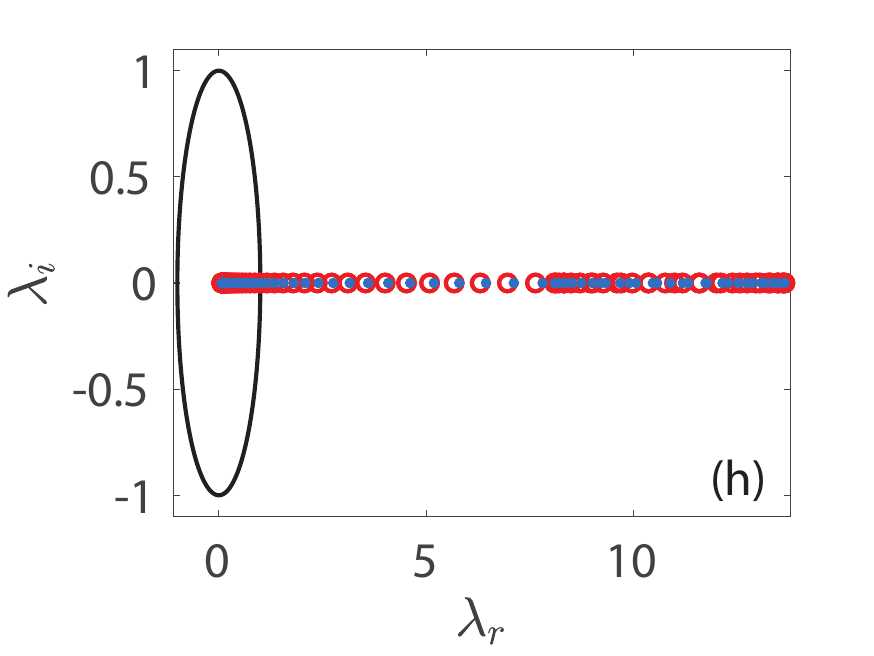} \quad
    \includegraphics[width=0.23\textwidth]{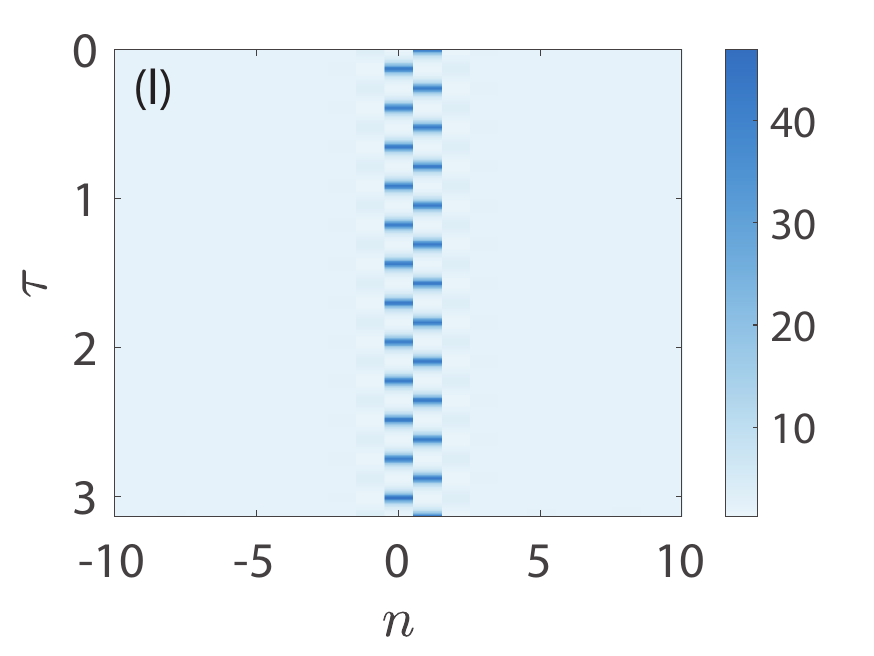}\quad
    \includegraphics[width=0.23\textwidth]{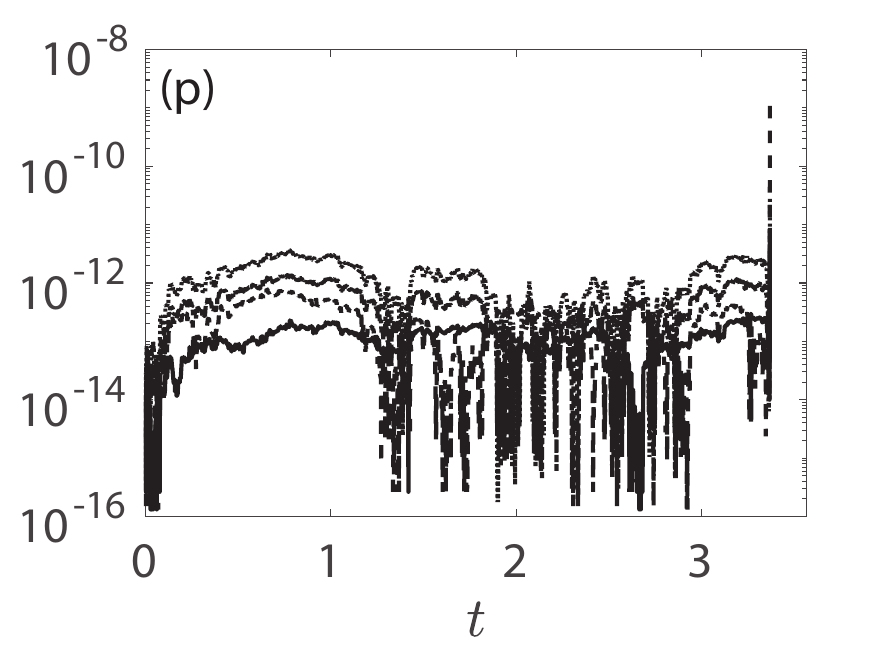}\quad \\
	\caption{
Summary of numerical results associated with the KM solutions of Fig.~\ref{1KM_Figure}(e),
(b), (c), and (d) shown here in the first, second, third, and fourth rows, respectively.
~The panels (a)-(d) show the spatial distribution of numerically exact KM
breathers with blue open circles (connected by solid line
as a guide to the eye).~For comparison, the exact solution is shown too with
red stars (see, the legend in panel (a)).
As expected, in all cases, the two are practically identical.
The spectral stability results are summarized
in panels (e)-(h) where the Floquet multipliers $\lambda=\lambda_{r}+\mathrm{i}\lambda_{i}$
are shown with blue filled circles against the MI prediction associated
with the uniform background state shown with red open circles.~Moreover,
the spatio-temporal evolution of $Q_{n}(\tau)$ is depicted in (i)-(l) where the respective
conserved quantities (in a semi-logarithmic scale) are shown in panels (m)-(p); see the
legend and Eqs.~\eqref{eq:consv_Gamma_1},~\eqref{eq:consv_Gamma_2},~\eqref{eq:consv_cinf},
and~\eqref{eq:consv_Ham}, as well as the associated discussion in the
text.
\label{1KM_num_results}
}
\end{figure}

We {now} turn our focus on Fig.~\ref{1KM_num_results} which presents numerical results on the existence,
stability and spatio-temporal dynamics in panels (a)-(d), (e)-(h), and (i)-(l), respectively.~Those are
complemented by the panels (m)-(p) which depict the relative error (with respect to $t=0$) of the conserved
quantities given by Eqs.~\eqref{eq:consv_Gamma_1},~\eqref{eq:consv_Gamma_2},~\eqref{eq:consv_cinf}, and~\eqref{eq:consv_Ham},
respectively (see the legend in panel (m)) all as a function of time.~Moreover, the first, second, third,
and fourth rows of Fig.~\ref{1KM_num_results} connect with the KM breathers of Figs.~\ref{1KM_Figure}(e),
(b), (c), and (d), respectively.~The panels (a)-(d) showcase the spatial dependence of the amplitude  $Q_{n}$ of
the numerically obtained solution (with blue open circles) against the respective exact solution (with red
stars), see the legend in panel (a).~Besides the expected matching between numerically exact and analytical
solutions, we report that the ones shown in panels (a) and (c) of Fig.~\ref{1KM_num_results} bear two-site excitations,
i.e., bond centered waveforms, whereas the ones of panels (b) and (d) correspond to site-centered solutions.

The respective results of the Floquet spectra of the solutions shown in (a)-(d) are presented in panels (e)-(h)
where the Floquet multipliers $\lambda=\lambda_{r}+\mathrm{i}\lambda_{i}$ are depicted with blue filled circles.%
~To understand the spectrum of the KM breathers, we
complement the relevant spectra by including in each panel the Floquet multipliers (see, the legend in
panel (e)) of the respective constant background (CW) solution in panels (e)-(h) with red open circles (see, the MI
dispersion relation in~\cite{SCCKK2020}, cf. Eqs.~(10) and~(14) therein with $g=1$ and $C=-1$).~In all cases that
we briefly report herein, all KM breathers are deemed spectrally unstable due to the presence of a band of
real Floquet multipliers (i.e., with $\lambda_{i}=0$) where $|\lambda|>1$.~However, the instabilities identified in
these panels emanate from the instability of the background, that is, the CW is modulationally unstable here since the
background amplitude in all cases we consider is greater than $1$, i.e., $Q_{0}>1$ [cf.~\cite{CS2024}]. Indeed, executing the relevant computations in larger
lattices, we have verified that the bands of associated instabilities become
more dense as is expected due to the excitation of additional unstable wavenumbers
therein.
We find no additional unstable modes getting generated by the mounting of KM breathers atop the CW waveform
at least for the cases we have studied numerically.
This is in line with earlier results in both continuum~\cite{karachalios}
and discrete~\cite{SCCKK2020} problems of the discrete nonlinear
Schr{\"o}dinger type.
It should be mentioned in passing that the breather itself slightly
perturbs the Floquet multipliers of the CW solution, see, indicatively, panels (g) and (h).~More importantly, we find
that the site-centered solutions of panels (b) and (d) as well as the bond-centered one of panel (a) are less unstable
compared to the bond-centered solution of panel (c) which itself involves a maximal growth rate of
$\mathcal{O}\left(10^{7}\right)$ and which is accordingly
expected to be manifested over very short time scales.~The stability trait of the KM breathers is reflected in the spatio-temporal
evolution of the amplitude $|Q_{n}(\tau)|$ in panels in (i)-(l).~Here, we use the numerically exact solution as an
initial conditions, and advance Eq.~\eqref{e:AL} forward in time until the integrator breaks down due to the (numerically induced) blow-up of
the solutions.

~In line with our spectral stability calculations shown in panels (e)-(h), we observe that the waveforms
of panels (i), (j), and (l) do breathe in time, and in a staggered or beating fashion without changing their structural
characteristics until modulational instability (MI) gets triggered due to perturbations.~For example, see the waveform
shown in Fig.~\ref{1KM_num_results}(i) close to the end of the time window, but also the one of Fig.~\ref{1KM_num_results}(k)
where MI gets manifested itself more quickly given the very large
instability growth rate, causing the solution to not even return back after one period of time
integration.

The perturbations mentioned above stem from not only round-off and local truncation errors of the
integrator, but also from the error that gets introduced by the truncation of Eq.~\eqref{eq:fourier_decomp},
i.e., in the computation of the initial condition as a fixed-point solution.~For reference, we
report the integration times $[0,t_{f}]$ of the respective cases with (i) $t_{f}=22.143T_{b}\approx 1.0322$,
(j) $t_{f}=18.4 T_{b}\approx 9.7596$, (k) $t_{f}=T_{b}\approx 1.6265$, and (l) $t_{f}=13T_{b}\approx 3.39720$,
respectively.~It should be noted that the spatio-temporal dynamics of $|Q_{n}(\tau)|$ shown in these panels
are up to times where the solution has not blown up yet, i.e., up to times slightly before blow-up where the
structures can still be visible.~Despite the evolution of the KM breathers towards blow-up, we report that the conserved quantities introduced in Eqs.~\eqref{eq:consv_Gamma_1},~\eqref{eq:consv_Gamma_2},~\eqref{eq:consv_cinf},
and~\eqref{eq:consv_Ham} are very well preserved with relative errors $\mathcal{O}(10^{-9})-\mathcal{O}(10^{-16})$.%
This happens even very close to blow-up times as this shown in panels (m)-(p) where the terminal times therein
are up to and including those when the integrator breaks down.~This is an additional indication of the fidelity of
our numerical computations discussed in this section.

We speculate that what is happening here is that the relevant
solutions are very proximal to ones featuring collapse dynamics,
as discussed in the previous sections, given the relevant variation of
parameters such as $c$. The dynamical manifestation of the waveforms'
MI (often with a very large growth rate)
due to its background rapidly leads to growth and opens a pathway
towards exploring these proximal collapse-prone dynamical features
of the model. Understanding these pathways towards collapse
is naturally an interesting direction for further exploration.
This is especially so since it's worthwhile to bear in mind that the
exact analytical solutions constructed herein have been designed
to feature no collapse, and it is only the quite small perturbations
of roundoff, integration local truncation error, and inexactness due to
a finite mode truncation that lead to the observed blowup.

\section{Concluding remarks}
\label{s:concl}

In the present work
we have examined the existence (including the analytical form),
stability and dynamics of an important, periodic-in-time class of solutions of the defocusing Ablowitz-Ladik lattice on an arbitrarily large background which are discrete analogs of the
KM breathers of the focusing NLS equation.
Specifically, we derived conditions on the background and on the spectral parameters that guarantee the exact KM solution to be non-singular on the lattice for all times. Furthermore, a novel KM-type breather solution is presented which is also regular on the lattice under suitable conditions. Furthermore, we employed Darboux transformations to obtain a multi-KM breather solution, and showed that parameters choices exist for which the
newly obtained double KM breather to be regular on the lattice.
Finally, we presented numerical results on the  stability and
spatio-temporal dynamics of the single KM breathers. The stability analysis
manifested that these states were modulationally unstable due to
the modulational instability of the background on top of which they
existed, with their Floquet multipliers suitably modified (due to
the nonlinear nature of the waveform), yet without novel instabilities
introduced (similarly to what was found for the continuum NLS problem
in~\cite{karachalios}). Another important feature of the numerical
computations was that small perturbations (introduced by roundoff
and local truncation errors, or by the approximate nature of our
numerical solution) led to the finite time blowup of the KM structures.
The preservation of the model's conservation laws strongly suggested
that this feature was not a numerical artifact, but rather a
real feature which merits further examination.

In a forthcoming work, a more systematic stability analysis will be carried out together with homotopic continuations of solutions presented herein towards the discrete nonlinear Schr\"odinger (DNLS) equation, and in the same spirit as in~\cite{SCCKK2020}. This may provide significant
insights towards the existence of corresponding KM waveforms in the
more experimentally relevant (within nonlinear optics) model of the
discrete NLS equation~\cite{dnlsbook}. Another interesting direction
is to continue the waveforms (especially the novel ones discovered
herein) towards smaller frequencies in order to acquire structures
closer to the rogue waves that are known to exist
in the AL model (see, e.g.,~\cite{AASC2010}).
Further such studies of rogue waves in both defocusing and
focusing AL lattices and their non-integrable analogs
are currently in progress and will be reported in future publications.


\section*{Acknowledgments}
This work has been supported by the U.S. National Science Foundation under Grants No. DMS-2204782 (EGC),
and DMS-2110030 and DMS-2204702 (PGK), and DMS-2406626 (BP).
The authors also acknowledge the Isaac Newton Institute for Mathematical Sciences, Cambridge, UK, for support and hospitality during the satellite programme ``Emergent phenomena in nonlinear dispersive waves'' (supported by EPSRC grant EP/R014604/1), where work on this paper was undertaken. Finally, we would also like to thank Gino Biondini for his suggestions with one of the figures, and the anonymous referee for their useful comments which helped us improve the work.

\appendix

\section{Construction of the Darboux matrix}

The DM in \cite{CS2024} is assumed to be of the form
\begin{equation}
\label{e:DMansatz}
D_n^{[0,J]}(z,\tau)=
\begin{pmatrix}
z^J+\sum_{l=1}^Ja_n^{(J-2l)}z^{J-2l} & \sum_{l=1}^Jb_n^{(J-2l+1)}z^{J-2l+1} \\
\sum_{l=1}^J \left(b_n^{(J-2l+1)}\right)^*z^{-J+2l-1} & z^{-J}+\sum_{l=1}^J \left(a_n^{(J-2l)}\right)^*z^{-J+2l}
\end{pmatrix}
\end{equation}
where the symmetry has already been taken into account. The unknown quantities in the DM are then determined
by requiring that the DM is singular at the discrete eigenvalues $\pm z_j,\pm 1/z_j^*$, with
\begin{equation}
\label{e:DMprop}
D_n^{[0,J]}(z_j,\tau)\phi_n^{[0]}(z_j,\tau)=\gamma_j D_n^{[0,J]}(z_j,\tau)\psi_n^{[0]}(z_j,\tau)\,,
\end{equation}
where $\Psi_n^{[0]}(z,\tau)=\left( \phi_n^{[0]}(z,\tau),\psi_n^{[0]}(z,\tau)\right)$ is a fundamental analytic matrix solution of the Lax pair
corresponding to the potential $Q_n^{[0]}(\tau)$, and $\gamma_j$ are the proportionality constants that relate the column vectors
at the discrete eigenvalue.
Inserting the ansatz \eqref{e:DMansatz} into \eqref{e:DMprop}, one obtains
\begin{gather}
\left( z_j^J+\sum_{l=1}^Ja_n^{(J-2l)}z_j^{J-2l} \right)+\left( z_j^J+\sum_{l=1}^J b_n^{(J-2l+1)}z_j^{J-2l+1} \right) r_j=0,
\\
\left( \frac{1}{(z_j^*)^J}\sum_{l=1}^J\frac{a_n^{(J-2l)}}{(z_j^*)^{J-2l}} \right)+\left( \sum_{l=1}^J \frac{b_n^{(J-2l+1)}}{(z_j^*)^{J-2l+1}}\right)\frac{1}{r_j^*}=0,
\end{gather}
where
$$
r_j=\frac{\phi_n^{[0],2}(z_j)-\gamma_j \psi_n^{[0],2}(z_j)}{\phi_n^{[0],1}(z_j)-\gamma_j \psi_n^{[0],1}(z_j)}
$$
and the superscript denotes the corresponding component of the vectors $\phi_n^{[0]}(z_j,\tau)$, $\psi_n^{[0]}(z_j,\tau)$.
If the number of singular points $z_j$ is equal to the order $J$ of the
Darboux transformation, the above equations for $j=1, \dots,J$  define an linear system of $2J$ equations for the $2J$ unknowns
$a_n^{(l)},b_n^{(l)}$ for $l=1,\dots,J$. Finally, the solution $Q_n^{[J]}(\tau)$ can be obtained from $Q_n^{[0]}(\tau)$ using the first of \eqref{e:DE}, yielding
\begin{equation}
\label{e:DTpot}
Q_n^{[0,J]}(\tau)=Q_n^{[0]}(\tau)a_{n+1}^{(-J)}(\tau)+b_{n+1}^{(-J+1)}(\tau).
\end{equation}

The solutions of the above equations for the simplest cases, $J=1$ and $J=2$, are provided in \cite{CS2024}. Specifically, if $J=1$:
\begin{gather}
a_n^{(-1)}(\tau)=\frac{z_1-|r_1|^2/z_1^*}{|r_1|^2z_1^*-1/z_1}\,, \qquad b_n^{(0)}(\tau)=-\frac{r_1^*}{|r_1|^2z_1^*-1/z_1}\left( |z_1|^2-1/|z_1|^2\right),
\end{gather}
yielding
\begin{equation}
Q_n^{[0,1]}(\tau)=Q_n^{[0]}(\tau)a_{n+1}^{(-1)}(\tau)+b_{n+1}^{(0)}(\tau)\,.
\end{equation}
Similarly, when $J=2$ the DM elements are given by:
\begin{equation}
a_n^{(-2)}(\tau)=\frac{\Delta_a^{(2)}}{\Delta^{(2)}}\,, \qquad b_n^{(-1)}(\tau)=\frac{\Delta_b^{(2)}}{\Delta^{(2)}}\,,
\end{equation}
where
\[
\Delta^{(2)}=\det\begin{pmatrix}
1 & 1/z_1^2 & r_1 z_1 & r_1/z_1 \\
1 & 1/z_2^2 & r_2z_2 & r_2/z_2 \\
1 & (z_1^*)^2 & 1/(r_1z_1)^* & z_1^*/r_1^* \\
1 & (z_2^*)^2 & 1/(r_2z_2)^* & z_2^*/r_2^*
\end{pmatrix}
\]
$\Delta_a^{(2)}$, $\Delta_b^{(2)}$ are obtained replacing the second and forth column, respectively, with the vector
$$
(-z_1^2,-z_2^2,-(z_1^*)^{-2}, -(z_2^*)^{-2})^T,
$$
and the potential is reconstructed by means of Eq.~\eqref{e:DTpot}, i.e.
\begin{equation}
Q_n^{[0,2]}(\tau)=Q_n^{[0]}(\tau)a_{n+1}^{(-2)}(\tau)+b_{n+1}^{(-1)}(\tau)\,.
\end{equation}

\section{Errata for Ref.~\cite{OP2019}}

\begin{enumerate}
\item The first equation without number in \textit{Remark 1} should read
$$
|\lambda|^2\le 1 \quad \Leftrightarrow \quad (|\zeta|^2+1)\left[|\zeta|^2+ir(\zeta-\zeta^*)-1\right]\le 0 \quad \Leftrightarrow \quad |\zeta-ir|^2\le Q_o^2
$$
\item The expression between Eq.~(32b) and Eq.~(33a) should read
    \begin{equation}
        (M_n(\tau,\zeta)\quad \bar{M}_n(\tau,\zeta))=(\bar{N}_n(\tau,\zeta)\quad N_n(\tau,\zeta))\Lambda^{n} e^{i\sigma_3\Omega{(\zeta)}\tau}\Tilde{T}(\zeta)e^{-i\sigma_3\Omega{(\zeta)}\tau}\Lambda^{-n}
    \end{equation}
   where $\Tilde{T}=A(\lambda(\zeta)) T(z(\zeta)) A^{-1}(\lambda(\zeta))$ is defined as in \cite{OP2019} as
\item As a result, Eqs.~(38c) and (38d) should respectively be
    \begin{equation}
        \beta (\zeta) = -\dfrac{Q_0^2\Delta_n(\tau)}{ir(\zeta +1/\zeta -2ir)}
        \left[\dfrac{\zeta(\zeta-ir)}{ir\zeta - 1}\right]^{n}
        e^{2i\Omega(\zeta)\tau}
        Wr(M_n(\tau,\zeta), \bar{N}_n(\tau,\zeta))
    \end{equation}
    \begin{equation}
        \bar{\beta} (\zeta) = \dfrac{Q_0^2\Delta_n(\tau)}{ir(\zeta +1/\zeta -2ir)}
        \left[\dfrac{\zeta(\zeta-ir)}{ir\zeta - 1}\right]^{-n}
        e^{-2i\Omega(\zeta)\tau}
        Wr(\bar{M}_n(\tau,\zeta), N_n(\tau,\zeta))
    \end{equation}
\item There is a typo in a sign in the exponents in Eqs.~(63a) and (63b). The correct expression for (63a) and (63b) is respectively shown below
    \begin{align}
        Res_{\zeta = \Bar{\zeta}_k} \frac{\Bar{M}_n(\tau,\zeta)}{\Bar{a}(\zeta)} &= \Bar{C}_k\lambda^{2n}(\Bar{\zeta}_k)e^{2i\Omega(\Bar{\zeta}_k)\tau}\bar{N}_n(\tau,\Bar{\zeta}_k), \qquad \Bar{C}_k = \frac{\Bar{b}_k}{\lambda(\Bar{\zeta}_k)\Bar{a}'(\Bar{\zeta}_k)} \\
        Res_{\zeta = {\zeta}_k} \frac{{M}_n(\tau,\zeta)}{{a}(\zeta)} &= {C}_k\lambda^{-2n}({\zeta}_k)e^{-2i\Omega({\zeta}_k)\tau}{N}_n(\tau,{\zeta}_k), \qquad \Bar{C}_k = \frac{{b}_k\lambda({\zeta}_k)}{{a}'({\zeta}_k)}
    \end{align}
    \item Equation (63b) is used to obtain equations (68a) and so it contains a typo in a sign in the exponent. The correct expression for (68a) is
    \begin{align}
        \Bar{N}_n(\tau,\zeta) =& \begin{pmatrix} 1/\Delta_n(\tau)\\
        (\zeta-ir)/Q_+
        \end{pmatrix} +\sum_{k=1}^{2J} \frac{(\zeta-ir)}{(\zeta_k-ir)(\zeta-\zeta_k)} {C}_k\lambda^{-2n}({\zeta}_k)e^{-2i\Omega({\zeta}_k)\tau}{N}_n(\tau,{\zeta}_k)\nonumber\\
        &-\frac{1}{2\pi i}\int_{w\in \Sigma}
        \frac{(\zeta-ir)}{(w-\zeta)(w-ir)} {C}_k\lambda^{-2n}({w})e^{-2i\Omega({w})\tau}{N}_n(\tau,w){\rho}(w)dw
    \end{align}
    \item Similarly, Equation (63a) is used to obtain equations (68b), whose correct expression is
    \begin{align}
        {N}_n(\tau,\zeta) =& \begin{pmatrix} (1/\zeta-ir)/Q_+^*\\
        1/\Delta_n(\tau) \end{pmatrix} +\sum_{k=1}^{2J} \frac{(\zeta+i/r)}{(\Bar{\zeta}_k+i/r)(\zeta-\Bar{\zeta}_k)} \Bar{C}_k\lambda^{2n}(\Bar{\zeta}_k)e^{2i\Omega(\Bar{\zeta}_k)\tau}\Bar{N}_n(\tau,\Bar{\zeta}_k)\nonumber\\
        &+\frac{1}{2\pi i}\int_{w\in \Sigma}
        \frac{(\zeta+i/r)}{(w-\zeta)(w+i/r)} \lambda^{2n}({w})e^{2i\Omega({w})\tau}\Bar{N}_n(\tau,w)\bar{\rho}(w)dw
    \end{align}
    \item Following the correction in (68b), Equations (69a) and (69b) are respectively  corrected as
    \begin{align}
        \frac{1}{\Delta_n(\tau)} =& 1 + Q_+^* \sum_{k=1}^{2J} \frac{1}{1-ir\Bar{\zeta}_k} \Bar{C}_k\lambda^{2n}(\Bar{\zeta}_k)e^{2i\Omega(\Bar{\zeta}_k)\tau}\Bar{N}_n^{(1)}(\tau,\Bar{\zeta}_k)\nonumber\\
        &-\frac{Q_+^*}{2\pi i}\int_{w\in \Sigma}
        \frac{1}{1-irw} \lambda^{2n}({w})e^{2i\Omega({w})\tau}\Bar{N}_n^{(1)}(\tau,w)\bar{\rho}(w)dw\\
        \frac{Q^*_n(\tau)}{\Delta_n(\tau)} =& \frac{Q^*(\tau)}{\Delta_n(\tau)} + Q_+^* \sum_{k=1}^{2J} \frac{1}{\Bar{\zeta}_k +i/r} \Bar{C}_k\lambda^{2n}(\Bar{\zeta}_k)e^{2i\Omega(\Bar{\zeta}_k)\tau}\Bar{N}_n^{(2)}(\tau,\Bar{\zeta}_k)\nonumber\\
        &-\frac{Q_+^*}{2\pi i}\int_{w\in \Sigma}
        \frac{1}{w+i/r} \lambda^{2n}({w})e^{2i\Omega({w})\tau}\Bar{N}_n^{(2)}(\tau,w)\bar{\rho}(w)dw
    \end{align}
    \item
    Consequently, equation (70) is corrected as
    \begin{equation}
        Q^*_n(\tau) = Q^*_+\left(1 +
        \frac{
        \sum_{k=1}^{2J}( \Bar{C}_k\lambda^{2n}(\Bar{\zeta}_k)e^{2i\Omega(\Bar{\zeta}_k)\tau}\Bar{N}_n^{(2)}(\tau,\Bar{\zeta}_k))/ ({\Bar{\zeta}_k +i/r})
        }{ 1+
        Q_+^* \sum_{k=1}^{2J} ( \Bar{C}_k\lambda^{2n}(\Bar{\zeta}_k)e^{2i\Omega(\Bar{\zeta}_k)\tau}\Bar{N}_n^{(1)}(\tau,\Bar{\zeta}_k))/({1-ir\Bar{\zeta}_k})
        }
        \right)
    \end{equation}
\item Eq.~(87) should read
\begin{equation}
\xi=n\log \rho +2(\Im \bar{\Omega}_1) \tau\,, \qquad \eta=n\phi +2(\Re \bar{\Omega}_1) \tau\,.
\end{equation}
    \item
    Equation (92) should read
    \begin{equation}
        \bar{\Omega}_1 = i\omega, \quad\quad \omega = 2rQ_0^2\frac{ (Q_0 + r\sin \alpha)\cos \alpha}{2Q_0(Q_0+r\sin \alpha)-1}.
    \end{equation}
\end{enumerate}

\end{document}